\pdfoutput=1
\documentclass[11pt]{article}

\usepackage{jcappub}
\usepackage{amsmath}
\usepackage{amssymb}
\usepackage{braket}

\usepackage{graphicx}
\usepackage{mathrsfs}
\usepackage{subcaption}
\usepackage{empheq}
\usepackage{appendix}
\usepackage{natbib}
\usepackage[dvipsnames]{xcolor}
\usepackage{float}
\usepackage{indentfirst}
\usepackage[format=plain,indention=.5cm,labelfont={sc},textfont={sl}]{caption}
\usepackage[top=1in, bottom=0.4in, left=0.65in, right=0.65in, includefoot]{geometry}
\usepackage{xfrac}
\usepackage{sectsty}
\sectionfont{\centering}
\allowdisplaybreaks

\title{Extended interactions in the Palatini--$R^2$ inflation}
\author{Angelos Lykkas,}
\author{Kyriakos Tamvakis}
\affiliation{Physics Department, University of Ioannina, GR--45110 Ioannina, Greece}
\emailAdd{a.lykkas@uoi.gr}
\emailAdd{tamvakis@uoi.gr}

\abstract{The Palatini formulation of the Starobinsky model does not yield a propagating scalaron that can assume the role of the inflaton field as in the conventional metric formulation. In the so-called Palatini--${R}^2$ models this role is assumed by a fundamental scalar field nonminimally coupled to gravity. In this article we consider an extension of the interactions of this field by introducing a field dependence to the $R^2$ coefficient in the form of logarithmic corrections. We examine the resulting predictions of the inflationary observables in the framework of slow-roll approximation and investigate the possible implications of the reheating process on these predictions. We find that the model predictions lie in the favoured region of the observations and that the assumed scalar field dependence of the $R^2$ coupling tends to decrease the value of the spectral index $n_s$ and potentially enhance the tensor-to-scalar ratio $r$, circumventing the highly suppressed values commonly cited for models considered in the same framework.}

\begin{document}

\maketitle

\section{Introduction}
Since its first iteration the proposed scenario of cosmological inflation~\cite{Guth1981, Linde1982, Albrecht1982, Sato1981, Linde1983a}, namely an exponential de Sitter-like expansion of the universe at the very early stages after its genesis, has become the cornerstone of our understanding of the early universe. Even though it was initially motivated by providing solutions to long-standing issues of the Big Bang cosmology, notably the flatness and the horizon problems, it is presently revered for its rather elegant way of addressing the origin of large scale structure formation that is observed today~\cite{Starobinsky1979, Mukhanov1981, Hawking1982, Hawking1983, Starobinsky1982, Guth1982}. The quantum treatment of the inflaton field driving the expansion predicts spatial inhomogeneities that grow in time and are detected today in the cosmic microwave background (CMB). The inflationary era is, in its simplest realisation, modeled by the evolution of a scalar degree of freedom that in principle can be a fundamental scalar field or a propagating scalar mode of gravity. The latter is realized in the Starobinsky model~\cite{Starobinsky1980}, featuring an extended gravitational sector that includes a quadratic Ricci scalar term $R^2$, leading to predictions in excellent agreement with observations. Another compelling idea, along the line of a fundamental scalar inflaton, is that of the Higgs inflation~\cite{Bezrukov2008, DeSimone2009, Bezrukov2009a}, where the role of the inflaton is assumed by the Higgs boson of the electroweak interactions nonminimally coupled through a $\xi|H|^2R$ term, providing a link between cosmological inflation and particle physics, and thus giving an explanation for the inflaton's origin. In contrast to the robust status of the Starobinsky model, Higgs inflation requires large values of the nonminimal coupling $\xi$ in order to be in contact with observations. Nevertheless, since the exact particle physics mechanism describing inflation is yet unknown, numerous phenomenological versions of these general models are being proposed with predictions compatible with present observational data~\cite{Akrami2018, Ade2018}.

Even though the required period of inflation seems to apply evergrowing restrictions to its allowed model space, not that much is known of the period following the end of inflation until the point of baryogenesis and Big Bang nucleosynthesis (BBN). Admittedly, similar to minimal inflation, there is a minimal scenario of reheating, namely the rethermalisation of the contents of the cold universe at the final stages of inflation, through oscillations of the inflaton condensate around its potential minimum. Then, these now massive inflaton states decay to Standard Model (SM) degrees of freedom. While the underlying mechanism of reheating is highly uncertain and its observational bounds are relaxed, it leads to several models existing of varying complexity; from simpler ones dealing with perturbative effects of the inflaton~\cite{Abbott1982, Dolgov1982, Albrecht1982b} to others that include nonperturbative dynamics~\cite{Felder1999} such as parametric resonance decay~\cite{Kofman1994, Traschen1990, Kofman1997} and tachyonic instabilities~\cite{Greene1997, Felder2001, Felder2001a, Shuhmaher2006, Dufaux2006}. However, in a similar manner to the inflationary period, the era of reheating can be parametrised by the number of $e$-foldings $N_\text{R}$ encoding its duration, its state parameter $w_\text{R}$ and its temperature $T_\text{R}$. It was shown~\cite{Dodelson2003, Liddle2003a, Dai2014a, Munoz2014, Gong2015, Cook2015a} that these parameters are interconnected with the inflationary ones, such as the spectral index $n_s$ and the $e$-folds $N$ produced during inflation. Therefore, despite its theoretical uncertainty the process of reheating can provide additional information and constraints on the allowed inflationary models.

The whole paradigm of inflation and reheating is manifestly interwoven with the gravitational degrees of freedom and therefore their parametrisation. The theory of General Relativity (GR), and much of our success in understanding the gravitational interaction, was established on spacetime $\mathcal{M}$ with a metric tensor $g$ as the dynamical variable, and the connection $\Gamma$ axiomatically identified with the Levi-Civita connection, i.e. the unique torsionless and metric-compatible connection. However, it was shown~\cite{Palatini:1919} that if one treats the connection and the metric as dynamical quantities independent of each other, variation of the standard Einstein-Hilbert action leads to the same equations of motion. The main difference is that the Levi-Civita connection is obtained on shell. This is often dubbed as the Palatini or first order formulation of gravity. The mere peculiarity of the Palatini variation was elevated to a fundamental question on the parametrisation of the gravitational degrees of freedom due to later attempts to canonically quantise GR via a Hamiltonian framework and the hope of realising it as a gauge theory much like Yang-Mills. Additionally, extensions of the gravitational sector in the form of gravity-matter nonminimal couplings and the inclusion of higher-order curvature invariants, has breathed new life into the Palatini formalism~\cite{Bauer2008, Bauer2011b, Tamanini2011, Bauer2011, Borowiec2012, Stachowski2017,  Rasanen2017, Tenkanen2017, Racioppi2017, Markkanen2018, Jaerv2018, Fu2017, Enckell2018, Kozak2019, Enckell2019, Antoniadis2018, Rasanen2019a, Rasanen2019, Almeida2019, Antoniadis2019, Jinno2019, Tenkanen2019a, Edery2019, Rubio2019, Tenkanen2019b, Bostan2020, Tenkanen2020d, Gialamas2020, Racioppi2020, Shaposhnikov2021, Tenkanen2020b, Tenkanen2020c, Shaposhnikov2020b, LloydStubbs2020, Antoniadis2020, Ghilencea2020a, Das2020, Jaerv2020, Gialamas2020b, Karam2020, McDonald2020, Laangvik2020, Ghilencea2020, Shaposhnikov2021a, Shaposhnikov2020c, Gialamas2020a, Verner2020, Bekov2020, Enckell2020, Dimopoulos2021, Karam2021a, Karam2021}. These types of modifications are prominent in most of the inflationary models discussed today and they lead to distinct predictions differentiating between the Palatini and the conventional metric formulations.

The paper is organised in the following way. In section~\ref{section2} we consider the action of a real scalar field $\phi(x)$ that is nonminimally coupled to the Starobinsky model, schematically through a coupling of the form $(1+f(\phi))R+\alpha(\phi)R^2$. In the Palatini framework the scalar degree of freedom of the Starobinsky model is not dynamical and we are effectively left with a single-field model accompanied by higher-order kinetic terms of the inflaton field $\phi(x)$. In section~\ref{section3}, under the assumption of slow-roll inflation, the predictions regarding inflationary observables are compared with their observed values coming from the Planck 2018 collaboration~\cite{Akrami2018} (also BICEP2~\cite{Ade2018}). Following that, in section~\ref{section4} we discuss the reheating phase of that same model. While refraining from addressing any particular mechanism of reheating, we connect the reheating parameters to inflationary ones and apply further constraints on the model parameters, briefly illustrating how such a connection is formed, based on~\cite{Liddle2003a, Dodelson2003, Dai2014a, Munoz2014, Cook2015a, Gong2015}. Finally, we present our conclusions in section~\ref{section5}.

\section{General Palatini framework}\label{section2}
We consider a fundamental scalar field $\phi(x)$ coupled to gravity nonminimally through its couplings to the Ricci scalar $R(g,\Gamma)$ allowing up to quadratic terms of it. A most general action would be
\begin{equation}
    \mathcal{S}=\int\!\mathrm{d}^4x\,\sqrt{-g}\left\{\,\frac{1}{2}\left(1+f(\phi)\right)R\,-\frac{1}{2}\left(\nabla\phi\right)^2\,-V(\phi)\,+\,\frac{1}{4}\alpha(\phi)R^2\,\right\},
\end{equation} 
in natural units $M_P\!=\!c\!=\!\hbar\!=\!1$. The nonminimal couplings $f(\phi)R$ and $\alpha(\phi)R^2$, even if absent at the classical level, are expected to be generated by quantum corrections of $\phi$ in a curved gravitational background (see e.g.~\cite{Parker2009}). An equivalent form of the above effective classical action can be written in terms of an auxiliary scalar $\chi$ by the addition of the Gaussian term $-\sfrac{1}{4}\int\!\mathrm{d}^4x\,\sqrt{-g}\,\alpha(\phi)\left(\chi\,-R\right)^2$. The scalar form of the action is
\begin{equation}
    \mathcal{S}=\int\!\mathrm{d}^4x\,\sqrt{-g}\left\{\frac{1}{2}\left(1+f(\phi)+\alpha(\phi)\chi\right)R\,-\frac{1}{2}\left(\nabla\phi\right)^2-V(\phi)\,-\frac{1}{4}\alpha(\phi)\chi^2\,\right\}\,.
\end{equation}
We shall consider this action as a starting point within the framework of the Palatini formalism assuming that the connection ${\Gamma^\rho}_{\mu\nu}$ and the metric tensor $g_{\mu\nu}$ have no a priori dependence on each other and is determined at the level of equations of motion. Thus, a Weyl rescaling of the metric
\begin{equation}
g_{ \mu\nu}(x)\,=\,\left(1+f+\alpha\chi\right)^{-1}\bar{g}_{ \mu\nu}(x)
\end{equation} 
leads to the Weyl-rescaled Einstein-frame action
\begin{equation}\label{Action:Weyl-rescaled}
    \mathcal{S}[\phi,\chi,\bar{g},\Gamma]=\int\!\mathrm{d}^4x\,\sqrt{-\bar{g}}\left\{\,\frac{1}{2}\,\bar{g}^{\mu\nu}R_{\mu\nu}(\Gamma)\,-\frac{1}{2}\frac{\left(\bar{\nabla}\phi\right)^2}{\left(1+f(\phi)+\alpha(\phi)\chi\right)}\,-\frac{\left(\,V(\phi)+\frac{1}{4}\alpha(\phi)\chi^2\right)}{\left(1+f(\phi)+\alpha(\phi)\chi\right)^{2}}\right\}\,.
\end{equation}
Note that in the Palatini framework terms including derivatives of $\phi$ and more importantly $\chi$ do not arise from the rescaling of the metric.

Variation of the action \eqref{Action:Weyl-rescaled} with respect to $\Gamma^\rho_{\mu\nu}$ gives the standard Levi-Civita expression as an equation of motion, namely
\begin{equation}\label{Levi}
\frac{\delta{\cal{S}}}{\delta\Gamma_{ \mu\nu}^{ \rho}}\,=\,0\,\,\Longrightarrow\,\,\Gamma_{ \mu\nu}^{ \rho}\,=\,\frac{1}{2}\bar{g}^{ \rho\sigma}\left(\partial_{ \mu}\bar{g}_{ \sigma\nu}+\partial_{ \nu}\bar{g}_{ \mu\sigma}\,-\partial_{ \sigma}\bar{g}_{ \mu\nu}\right)\,.
\end{equation}
Variation with respect to the auxiliary $\chi$ gives rise to the constraint equation:
\begin{equation}
\frac{\delta{\cal{S}}}{\delta\chi}\,=\,0\,\,\Longrightarrow\,\chi\,=\,\frac{4V(\phi)+\left(1+f(\phi)\right)\left(\bar{\nabla}\phi\right)^2}{\left(1+f(\phi)\right)-\alpha(\phi)\left(\bar{\nabla}\phi\right)^2}\,.
\end{equation}
Substituting this expression back into the action \eqref{Action:Weyl-rescaled} we obtain
\begin{equation}\label{Action:Final}
\mathcal{S}[\bar{g},\phi]=\int\!\mathrm{d}^4x\,\sqrt{-\bar{g}}\left\{\frac{1}{2}\bar{R}\,-\frac{1}{2}K(\phi)\left(\bar{\nabla}\phi\right)^2\,+\,\frac{1}{4}W(\phi)\left(\bar{\nabla}\phi\right)^4-U(\phi)\right\}\,,
\end{equation}
where the noncanonical kinetic functions $K(\phi)$ and $W(\phi)$ together with the Einstein-frame scalar potential are defined as follows
\begin{align}
    K(\phi)&=\frac{1+f(\phi)}{\left[\left(1+f(\phi)\right)^2+4\alpha(\phi)V(\phi)\right]}\,,\label{Function-K}\\
    W(\phi)&=\frac{\alpha(\phi)}{\left[\left(1+f(\phi)\right)^2+4\alpha(\phi)V(\phi)\right]}\,,\label{Function-W}\\
    U(\phi)&=\frac{V(\phi)}{\left[\left(1+f(\phi)\right)^2+4\alpha(\phi)V(\phi)\right]}\,.\label{Potential-U}
\end{align}
Then, varying action \eqref{Action:Final} with respect to $\bar{g}_{ \mu\nu}$ we obtain the Einstein equation
\begin{align}\label{Eq:Gen-Einstein}
\overline{G}_{ \mu\nu}\equiv\overline{R}_{ \mu\nu}-\frac{1}{2}\overline{g}_{ \mu\nu}\overline{R}=&\left(K(\phi)-W(\phi)\left(\bar{\nabla}\phi\right)^2\right)\partial_{ \mu}\phi\,\partial_{ \nu}\phi\,-\bar{g}_{ \mu\nu}\left(\frac{1}{2}K(\phi)\left(\bar{\nabla}\phi\right)^2-\frac{1}{4}W(\phi)\left(\bar{\nabla}\phi\right)^4+U(\phi)\right)\,,
\end{align}
where, henceforth, the prime denotes derivative with respect to the function's argument. Finally, variation of eq.~\eqref{Action:Final} with respect to $\phi(x)$ gives the scalar field equation of motion
\begin{align}\label{Eq:Gen-KG}
\left(K(\phi)-W(\phi)\left(\bar{\nabla}\phi\right)^2\right)\overline{\square}\, \phi-W(\phi)\left(\partial_{ \mu}\left(\bar{\nabla}\phi\right)^2\right)\bar{g}^{ \mu\nu}\partial_{ \nu}\phi+\frac{1}{2}K'(\phi)\left(\bar{\nabla}\phi\right)^2-\frac{3}{4}W'(\phi)\left(\bar{\nabla}\phi\right)^4-U'(\phi)=0\,.
\end{align}
Therefore, assuming the Palatini variation of action \eqref{Action:Final} the resulting equations of motion governing the dynamical variables of the theory are given in eq.~\eqref{Eq:Gen-Einstein}, eq.~\eqref{Eq:Gen-KG} and the on-shell Levi-Civita condition \eqref{Levi}.

\section{Slow-roll Inflation}\label{section3}

For a spatially homogeneous scalar field $\phi(t)$ in a flat FRW background
\begin{equation}
\mathrm{d}s^2\,=\,-\mathrm{d}t^2\,+\,a^2(t)\,\delta_{ij}\,\mathrm{d}x^i\,\mathrm{d}x^j\,\Longrightarrow\,\,\bar{g}_{ \mu\nu}\,=\,\left(\begin{array}{cc}
-1\,&\,0\\
\,&\,\\
0\,&\,a^2(t)\,\delta_{ij}
\end{array}\right)\,,
\end{equation}
with $i,j,\ldots=1,2,3$ being the designated spatial indices, the equations of motion reduce to the generalised Friedmann equation
\begin{equation}
3\left(\frac{\dot{a}}{a}\right)^2\,=\,3H^2\,=\,\frac{1}{2}K(\phi)\,\dot{\phi}^2+\frac{3}{4}W(\phi)\,\dot{\phi}^4+U(\phi)\,=\,\rho
\end{equation}
and the scalar field equation \eqref{Eq:Gen-KG} becomes:
\begin{equation}
\left(K(\phi)+3W(\phi)\,\dot{\phi}^2\right)\ddot{\phi}+3H\left(K(\phi)+W(\phi)\dot{\phi}^2\right)\dot{\phi}+\frac{1}{2}K'(\phi)\,\dot{\phi}^2+\frac{3}{4}W'(\phi)\,\dot{\phi}^4+U'(\phi)\,=\,0\,.
\end{equation}
Focusing on slow-roll inflation amounts to assuming that the terms in the vacuum energy obey
\begin{equation}\label{Condition-SR1}
\frac{3}{4}\,W(\phi)\,\dot{\phi}^4\ll \frac{1}{2}\,K(\phi)\,\dot{\phi}^2\ll U(\phi)
\end{equation}
and that in the scalar field equation of motion we also assume
\begin{equation}\label{Condition-SR2}
|\ddot{\phi}|\ll |3H\dot{\phi}|\,.
\end{equation}
Eq.~\eqref{Condition-SR1} is equivalent to having started with the approximate matter action
\begin{equation}
    \mathcal{S}_{m}=\int\!\mathrm{d}^4x\,\sqrt{-\bar{g}}\left\{\,-\frac{1}{2}K(\phi)\left(\bar{\nabla}\phi\right)^2\,-U(\phi)\right\}\,.
\end{equation}
This action can be rewritten in a canonical form
\begin{equation}
    \mathcal{S}_m=\int\!\mathrm{d}^4x\,\sqrt{-\bar{g}}\left\{-\frac{1}{2}\left(\bar{\nabla}\Phi\right)^2\,-U(\phi(\Phi))\,\right\}
\end{equation}
in terms of a canonically normalised scalar field $\Phi$ defined through
\begin{equation}
    \Phi=\pm\int\!\mathrm{d}\phi\,\sqrt{K(\phi)}\,.
\end{equation}
The Potential Slow-Roll Parameters (PSR) describing the underlying dynamics of (slow-roll) inflation are defined as
\begin{align}
    \epsilon_V&=\frac{1}{2}\left(\frac{U'(\Phi)}{U(\Phi)}\right)^2\,=\,\frac{1}{2K(\phi)}\left(\frac{{U}'(\phi)}{U(\phi)}\right)^2\,,\\
    \eta_V&=\frac{U''(\Phi)}{U(\Phi)}\,=\,\frac{1}{K(\phi)}\left(\frac{{U}''(\phi)}{U(\phi)}\right)-\frac{1}{2}\frac{{K}'(\phi)}{K^2(\phi)}\left(\frac{{U}'(\phi)}{U(\phi)}\right)\,.
\end{align}
The duration of inflation, estimated in number of $e$-foldings, is given by the expression
\begin{equation}
N\,=\,\int_{\Phi_*}^{\Phi_f}\!\mathrm{d}\Phi\left(\frac{U(\Phi)}{U'(\Phi)}\right)\,=\,\int_{\phi_*}^{\phi_f}\!\mathrm{d}\phi\,K(\phi)\left(\frac{U(\phi)}{U'(\phi)}\right)\,,
\end{equation}
where $\Phi_*$ (or $\phi_*$) is the field value at the start of inflation and $\Phi_f$ (or $\phi_f$) is the field value at the end of inflation, defined by the condition $\epsilon_V(\phi_f)\equiv 1$. The above PSR parameters are related to the observable quantities in the CMB, namely the scalar spectral index (or primordial tilt) $n_s$, the tensor-to-scalar ratio $r$  and the power spectrum of scalar perturbations $\mathcal{A}_s$, as follows~\cite{Lyth1999}
\begin{align}
    n_s&=1-6\epsilon_V(\phi_*)\,+\,2\eta_V(\phi_*)\,,\label{ns}\\
    r&=16\epsilon_V(\phi_*)\,,\label{r}\\
    \mathcal{A}_s&=\frac{U(\phi_*)}{24\pi^2\epsilon_V(\phi_*)}\label{As}\,.
\end{align}
The quantities in eqs.\eqref{ns}-\eqref{As} are calculated at the horizon exit (where $\phi=\phi_*$) and are expressed up to first order in slow-roll parameters. Their observed values coming from the Planck collaboration~\cite{Akrami2018}. (similarly from BICEP2~\cite{Ade2018}) are
\begin{equation}\label{Condition:Planck}
    n_s=\left\{\begin{matrix}(0.9607,0.9691),&1\sigma\text{ region}\\ &\\ (0.9565,0.9733),&2\sigma\text{ region}\end{matrix}\right.\,,\qquad r\lesssim 0.056\,,\qquad \mathcal{A}_s\approx2.1\times 10^{-9}\,.
\end{equation}

\subsection{Specifying the nonminimal couplings} 

The general setup described in the previous sections is expressed in terms of a scalar potential $V(\phi)$ and two nonminimal coupling functions $f(\phi)$ and $\alpha(\phi)$. The nonminimal coupling of the scalar field to the Ricci curvature scalar, being dimensionful, will in general correspond to a polynomial in $\phi$, or $\phi^2$ if we are to respect an additional internal $\mathbb{Z}_2$ symmetry $\phi\mapsto-\phi$ for the rest of the action. We shall adopt the simplest choice, assuming
\begin{equation}\label{Coupling-Xi}
f(\phi)=\xi\,\phi^2\,,
\end{equation}
parametrising this coupling in terms of the dimensionless parameter $\xi$.

As far as the scalar potential is concerned we presume that only renormalisable self-interaction terms of $\phi$ will play a role and therefore consider a quartic self-interaction scalar potential of the form
\begin{equation}\label{Potential-V}
V(\phi)\,=\,\frac{\lambda}{4!}\phi^4\,,
\end{equation}
possibly corrected in the UV completion of the present classical setup by renormalisable corrections in the form of logarithmic corrections as $\phi^4\ln(\phi^2/\mu^2)$, with $\mu$ denoting the renormalisation scale. At this point, we are safe to assume that $\phi$ could be describing the Higgs scalar far away from its vacuum expectation value (vev). In what follows we consider values of the free parameters that are phenomenologically consistent with the Higgs field, while allowing for the possibility that $\phi(x)$ represents another additional field to the SM. Therefore, we assume values of the self-coupling constant $\lambda\approx\left.\lambda_H\right|_{\Lambda=M_P}\lesssim\mathcal{O}(10^{-4})$ and we suppose that new physics phenomena stabilise the Higgs self-coupling, which via conventional SM wisdom suggests that it runs to negative values within some margin of error.
 
The remaining parametric function $\alpha(\phi)$ corresponds to a generalisation of the Starobinsky constant.\footnote{See, e.g. \cite{Gundhi2021, Gundhi2021a} for a similar discussion in the metric formulation.} In contrast to the other nonminimal coupling $f(\phi)$, which could be transformed by a Weyl transformation, $\alpha(\phi)$ is Weyl-invariant.\footnote{The term $\sqrt{-g}\,g^{\mu\nu}g^{\rho\sigma}$ in $\alpha\sqrt{-g}\,R^2=\alpha\sqrt{-g}\left(g^{\mu\nu}R_{\mu\nu}(\Gamma)\right)^2$ is invariant under a Weyl rescaling of the metric in the Palatini formalism.} Therefore, a direct relation between these couplings would be entirely ad hoc, however, in contrast see~\cite{Das2020}. In an assumed UV completion of the theory $\alpha(\phi)$, being dimensionless, would be expected to acquire logarithmic corrections in the form
\begin{equation}\label{Coupling-Alpha}
    \alpha(\phi)\,=\,\alpha_0\,+\,\beta_0\ln(\phi^2/\mu^2)\,,
\end{equation}
where $\alpha_0,\,\beta_0$ are constant parameters. In what follows the value of the free parameter $\beta_0$ is accompanied by a factor of at least $\alpha_0/\beta_0\propto\mathcal{O}(10)$ in order to maintain the perturbative nature of the coupling $\alpha(\phi)$ and insure the positivity of the overall $R^2$ coefficient.

As the scalar field tends to values close to the renormalisation scale $\mu$, the model exhibits an asymptotic scale invariance and the original Starobinsky model is recovered with small corrections of the form $\sim\beta_0(\delta\phi/\mu)^2$, where $\delta\phi$ are pertubations around $\mu$.  Note that in the phenomenology of Palatini inflation it has been established that observables such as $\mathcal{A}_s$ and $n_s$ do not depend on constant $\alpha$~\cite{Enckell2019, Antoniadis2018} and therefore such newfound $\phi$-dependent corrections are not expected to modify this drastically. In contrast, both the number of $e$-folds and the tensor-to-scalar ratio depend on $\alpha$ and thus its field dependence is expected to lead to deviations from previous results.

We proceed by considering the above parametric functions \eqref{Coupling-Xi}-\eqref{Coupling-Alpha}. The functions $K(\phi),\,W(\phi)$ and $U(\phi)$, defined in eqs.~\eqref{Function-K}-\eqref{Potential-U} and appearing in the final action become:
\begin{align}
    K(\phi)&=\frac{1+\xi\phi^2}{(1+\xi\phi^2)^2+\displaystyle{\frac{\lambda}{6}}\phi^4\left(\alpha_0+\beta_0\ln(\phi^2/\mu^2)\right)}\,,\\
    W(\phi)&=\frac{\alpha_0+\beta_0\ln(\phi^2/\mu^2)}{(1+\xi\phi^2)^2+\displaystyle{\frac{\lambda}{6}}\phi^4\left(\alpha_0+\beta_0\ln(\phi^2/\mu^2)\right)}\,,\\
    U(\phi)&=\frac{\displaystyle{\frac{\lambda}{4!}}\phi^4}{(1+\xi\phi^2)^2+\displaystyle{\frac{\lambda}{6}}\phi^4\left(\alpha_0+\beta_0\ln(\phi^2/\mu^2)\right)}\,.\label{Eq:PotOr}
\end{align}

Note that we have not introduced explicitly logarithmic corrections of the scalar coupling. Such leading logarithmic corrections in the denominators of the kinetic functions and the potential can be readily incorporated in a redefinition of the $\alpha_0$ and $\beta_0$ parameters.\footnote{That is done through $\lambda(\phi)\alpha(\phi)\propto\lambda_0\alpha_0+\left(\lambda_1\alpha_0+\lambda_0\beta_1\right)\ln(\phi^2/\mu^2)$.} Their effect in the numerator of $U(\phi)$, apart from improving the flatness of the inflationary plateau in the intermediate region, is subleading due to the smallness of $\lambda$. Additionally, since the potential enters in the slow-roll parameters and therefore in the observables through its derivatives $U'/U,\,(U'/U)'$, $\lambda$-logarithmic corrections give rise to inverse scalar field powers which are subleading in the large field domain. As long as $\beta_0\ln(\phi^2/\mu^2)$ remains perturbative, the celebrated potential plateau of the Palatini framework stays practically unaffected, violated only logarithmically at Planckian scales $\phi \gtrsim \mu\sim \mathcal{O}(1)$, namely
\begin{equation}
 U(\phi)\,\approx\,\frac{\lambda}{4!\,\xi^2+4\lambda\left(\alpha_0+\beta_0\ln(\phi^2/\mu^2)\right)}\,.
\end{equation}
Expressing the above asymptotic potential in terms of the canonical field
\begin{equation}
\Phi=\int\!\mathrm{d}\phi\,\sqrt{K(\phi)}\,\stackrel{\xi\sqrt{\phi}\gg M_P}{\approx}\,\int\frac{\mathrm{d}\ln(\phi/\mu)}{\sqrt{\xi+\displaystyle{\frac{\lambda}{6\xi}}\left(\alpha_0+2\beta_0\ln(\phi/\mu)\right)}}\,=\,\frac{6\xi}{\lambda\beta_0}\sqrt{\xi+\frac{\lambda}{6\xi}\left(\alpha_0+2\beta_0\ln(\phi/\mu)\right)}\,,
\end{equation}
we obtain
\begin{equation}
U(\Phi)\,\approx\,\frac{3\xi}{2\lambda\beta_0^2}\left(\frac{1}{\Phi^2+\ldots}\right)\,,
\end{equation}
 where the dots signify exponentially small corrections of $\mathcal{O}\left(e^{-(\lambda\beta_0/6\xi)\Phi^2}\right)$.
 
\begin{figure}[H]
    \centering
    \includegraphics[width=0.65\textwidth]{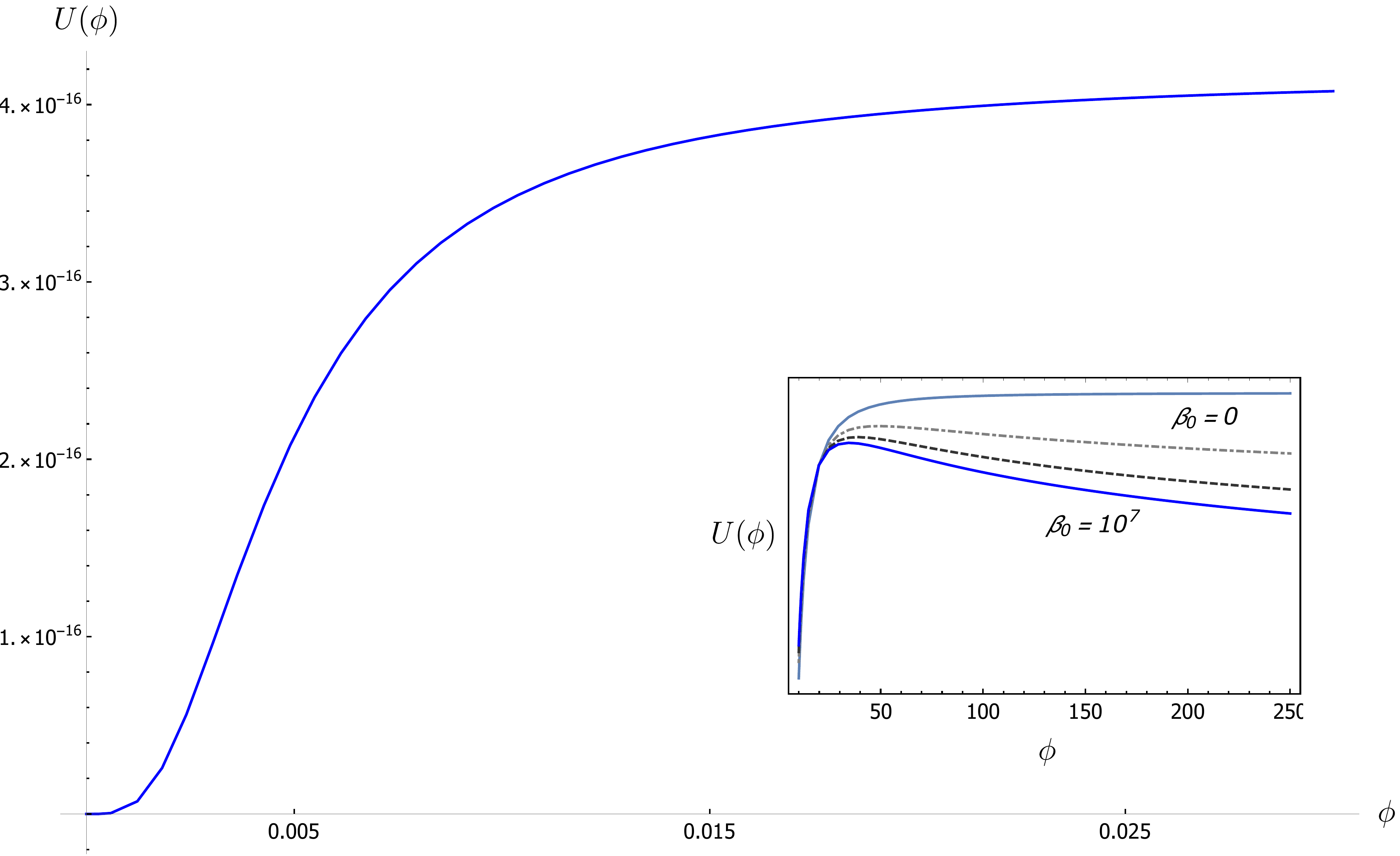}
    \caption{Plot of the scalar potential $U(\phi)$ in terms of the original field $\phi$, as given in eq.~\eqref{Eq:PotOr}. In the main figure the values of the free parameters are $\left\{\xi=10^5,\lambda=10^{-4},\mu=20,\alpha_0=10^8,\beta_0=10^7\right\}$. In the secondary figure we showcase the asymptotic behaviour of the potential in the large field limit for varying values of $\beta_0$, from $\beta_0\approx10^7$ to $\beta_0=0$. The local maximum of the potential is not relevant to inflation, since, as we show later on, inflation occurs for field values way before the maximum. It is also important to note that for values above the scale $\mu$, which is always assumed to $\simeq\phi_*$, the form of the potential cannot be trusted.}
    \label{figpot}
\end{figure}

 \subsection{Semi-analytic approach and numerical results}
 Considering the specified parametric functions \eqref{Coupling-Alpha} we obtain for the potential slow roll parameters $\epsilon_V$ and $\eta_V$
 \begin{align}
     \epsilon_V(\phi)&=\frac{1}{2\phi^2(1+\xi\phi^2)}\frac{\left(4(1+\xi\phi^2)-\frac{\lambda\beta_0}{3}\phi^4\right)^2}{\left(\,(1+\xi\phi^2)^2+\frac{\lambda}{6}\phi^4\left(\alpha_0+\beta_0\ln(\phi^2/\mu^2)\,\right)\,\right)}\,,\\
     \eta_V(\phi)&=3\,\epsilon_V(\phi)-\left(\frac{8}{\phi^2}+\frac{4}{\phi^2(1+\xi\phi^2)}+\frac{\lambda\beta_0}{3}\frac{\phi^2}{(1+\xi\phi^2)^2}\right)\,.
 \end{align}
 The corresponding expressions for the inflationary observable quantities are
 \begin{align}
     \mathcal{A}_s&\approx\frac{U(\phi)}{24\pi^2\epsilon_V}\,=\,\frac{1}{12\pi^2}\frac{\frac{\lambda}{4!}\phi^6(1+\xi\phi^2)}{\left(4(1+\xi\phi^2)-\frac{\lambda\beta_0}{3}\phi^4\right)^2}\,\\
     n_s&=1-\left(\frac{16}{\phi^2}+\frac{8}{\phi^2(1+\xi\phi^2)}+\frac{2\lambda\beta_0}{3}\frac{\phi^2}{(1+\xi\phi^2)^2}\right)\label{Equation:ns}\,\\
     r&\approx 16\epsilon_V=
 \frac{8}{\phi^2(1+\xi\phi^2)}\frac{\left(4(1+\xi\phi^2)-\frac{\lambda\beta_0}{3}\phi^4\right)^2}{\left[\,(1+\xi\phi^2)^2+\frac{\lambda}{6}\phi^4\left(\alpha_0+\beta_0\ln(\phi^2/\mu^2)\,\right)\,\right]}\,.\label{Eq:r}
 \end{align}
 Substituting the exact expressions of $K(\phi)$ and $U(\phi)$ in the integral formula for the number of $e$-foldings we obtain the following exact expression
 \begin{equation}\label{Equation:Ninfl}
     N=\left.\frac{b}{8\sqrt{4b+(\xi b)^2}}\left\{\left(1+\frac{1}{2}\xi^2b\right)\ln\left|\frac{\phi^2-\frac{1}{2}\xi b-\frac{1}{2}\sqrt{4b+(\xi b)^2}}{\phi^2-\frac{1}{2}\xi b+\frac{1}{2}\sqrt{4b+(\xi b)^2}}\right|\,+\,\frac{1}{2}\xi\sqrt{4b+(\xi b)^2}\ln\left|\phi^4-\xi b\phi^2-b\right|\,\right\}\right|_{\phi_f}^{\phi_*}
 \end{equation}
 where $b\equiv 12/(\lambda\beta_0)$. 
 
 Before we proceed to discuss the general case we note that in the minimal coupling case of $\xi=0$ the required order of magnitude for $\mathcal{A}_s$ implies $\lambda{\beta_0}^2\phi^2\sim\mathcal{O}(10^6)$, something that can only be met with a very large value for $\beta_0$.\footnote{In the previously studied minimal case, where $\xi\!=\!0\!=\!\beta_0$, the model describes successfully the inflationary era with appropriate inflationary observables, albeit requiring a larger than usual number of $e$-foldings $N\approx75$ $e$-folds~\cite{Antoniadis2019, Tenkanen2019a}. If now $\beta\neq0$ we expect large values of $\beta_0$ (not unlike the ones cited for $\alpha_0$ in the previous case~\cite{Tenkanen2019a, Tenkanen2020b}), since $\lambda_\text{max}\sim\mathcal{O}(10^{-4})$ and $\phi$ is around the Planck scale.} However, unreasonably large values of $\beta$ would invalidate our slow-roll assumption concerning the subleading nature of the quartic kinetic terms, expressed by eq.~\eqref{Condition-SR1}, which in the minimal case reads as
 $ 3\,\displaystyle{\frac{W(\phi)}{K^2(\phi)}}U(\phi)\ll 1$ or $\lambda\phi^4\alpha(\phi)\ll 1$.  In addition to that, a large value for $\beta_0$ and therefore $\alpha_0$ would result in a value of the effective speed of sound
 $c_s^2=\left(K(\phi)+W(\phi)\dot{\phi}^2\right)/\left(K(\phi)+3W(\phi)\dot{\phi}^2\right)$ significantly different than unity, which can also reach negative values indicating possible instabilities or exotic/unphysical states.
 
 Next, let us consider values for the parameters that can roughly satisfy the obsevables, starting from 
\begin{equation}
    \phi_*\sim 20 \qquad \text{and} \qquad \lambda\beta_0/\xi\sim\mathcal{O}(10^{-9})\,.
\end{equation}
The values presented above are of no particular interest and they are chosen (their ratio) approximately in order to satisfy the observational bounds of the inflationary quantities.  We remind the reader that the values of $\alpha_0$ and $\beta_0$ considered here are characterised by $\alpha_0/\beta_0\propto\mathcal{O}(10)$ based on perturbativity grounds. The ensuing discussion also shows that this phenomenologically based constrain does not affect the predictions of the inflationary observables. The assumed field value of $\phi_*$ is intimately tied to values of the spectral index $n_s$. Clearly, from eq.~\eqref{Equation:ns} the first order correction is $\propto\phi^{-2}$ with subleading corrections $\propto\left(\lambda\beta_0/\xi\right)\phi^{-4}$. Due to the smallness of these couplings it is safe to assume that $\phi_*\sim20$ and satisfy the observational bounds of $n_s$. Collectively, the predictions of the inflationary observables with respect to those values are 
\begin{equation}
    n_s\approx0.960\,,\qquad \mathcal{A}_s\sim 3.5\times10^{-9}\,,\qquad r\approx \frac{10^{-4}}{2\xi}
\end{equation}
and the corresponding number of $e$-folds can be obtained from the approximate formula
\begin{equation}
    N\approx\frac{3\xi}{2\lambda\beta_0}\ln\left|\frac{1-\displaystyle{\frac{\lambda\beta_0}{12\xi}}\phi_*^2}{1-\displaystyle{\frac{\lambda\beta_0}{12\xi}}\phi_f^2}\right|\,,
\end{equation}
resulting in $N\sim 50$ $e$-folds with $\phi_f^2\approx\mathcal{O}(\sqrt{8/\xi})$.
 
Some characteristic values for the parameters and the observables are shown in the table below. Entries of table~\ref{table1} are produced by first obtaining the field value $\phi_f$ at the end of inflation through the equation $\epsilon_V(\phi_f)\equiv 1$. Then, solving eq.~\eqref{Equation:Ninfl} for some specific value of $N\in[50,60]$, the field value $\phi_*$ at the start of inflation is readily obtained. Finally, we obtain the observables $r$ and $n_s$ while demanding that the parameter values also lead to the desired amplitude $\mathcal{A}_s$, as presented in eq.~\eqref{Condition:Planck}.

\begin{table}[h]
\centering
\begin{tabular}{||c|c||c|c||c|c||}
\hline
    $\xi$ & $\lambda$ & $r\,(N=50)$ & $r\,(N=60)$ &  $n_s\,(N=50)$ & $n_s\,(N=60)$  \\\hline\hline
    $10^{5}$&$10^{-4}$&$8\times10^{-9}$&$5.5\times10^{-9}$&$0.9600$&$0.9667$\\
    $10^{3}$&$10^{-6}$&$8\times10^{-8}$&$5.5\times10^{-8}$&$0.9600$&$0.9667$\\
    $10^3$&$10^{-5}$&$8\times10^{-7}$&$5.5\times10^{-7}$&$0.9600$&$0.9667$\\
    $10^2$&$10^{-7}$&$8\times10^{-6}$&$5.5\times10^{-6}$&$0.9600$&$0.9667$\\
    \hline
    \end{tabular}
    \caption{A numerical study of the exact expressions regarding inflationary observables. Here, we assumed constant values of $\alpha_0=10$, $\beta_0=1$ and $\mu=20\,M_P\sim\phi_*$ that, together with $\xi$ and $\lambda$, reproduce the appropriate value of the scalar amplitude $\mathcal{A}_s$. The field excursion $\Delta\phi\equiv\phi_*-\phi_f$ that is presented in the table, is approximated by $\phi_f\approx 10^{-1}M_P$ and $\phi_*\sim20M_P$.}
    \label{table1}
\end{table}

It turns out that different values of $\left\{\xi,\,\lambda,\,\alpha_0,\,\beta_0\right\}$ discussed above, lead to absolutely identical values of $n_s$. This is evident directly from eq.~\eqref{Equation:ns} and it is a feature of the Palatini--$R^2$ models (see e.g.~\cite{Antoniadis2018, Antoniadis2019, Gialamas2020}). It is also worth mentioning that the tensor-to-scalar ratio $r$ is highly suppressed and effectively undetectable, as is presented in a more comprehensive manner in the following figure.

\begin{figure}[H]
    \centering
    \includegraphics[scale=0.48]{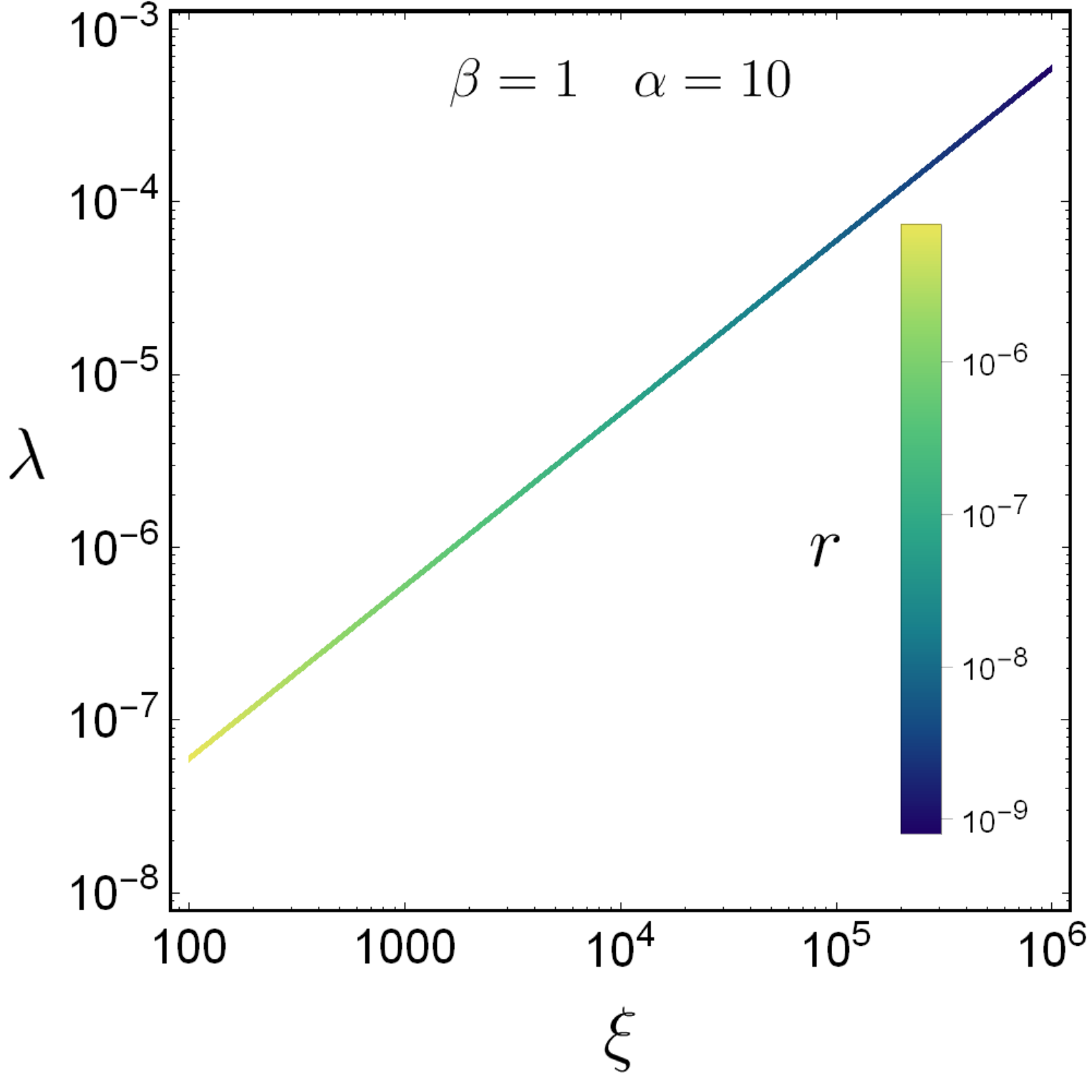}
    \caption{Plot of $\lambda=f(\xi)$ and the associated values of $r$ corresponding to the color grading displayed in the bar of the figure. The values presented are calculated at $N=50$ $e$-folds and the value of $\mathcal{A}_s\approx2.1\times10^{-9}$ is maintained throughout the curve. Displayed in the figure is purely the part of a larger contour plot, that leads to the correct value of the scalar power spectrum $\mathcal{A}_s$.}
    \label{fig1}
\end{figure}

In the limit of $\beta_0\to0$ it was demonstrated~\cite{Gialamas2020, Antoniadis2020, Gialamas2020b, Tenkanen2019a, Tenkanen2020c} that in order for the Palatini--$R^2$ models to be in contact with observations (namely the tensor-to-scalar ratio $r$) it requires values of $\alpha_0\sim10^8$. Similar values are reported in ref.~\cite{LloydStubbs2020} in order to retain the canonically normalised inflaton field $\Phi$ at sub-Planckian values. So, to make contact with previous works we allow for larger values of $\alpha_0$ and therefore $\beta_0$, while keeping the ratio of $\alpha_0/\beta_0\sim\mathcal{O}(10)$. As is directly noticeable from eq.~\eqref{Equation:ns} and eq.~\eqref{Eq:r}, larger values of $\beta_0$ can potentially impact the values of the inflationary observables substantially, albeit in a negative way.

\begin{table}[h]
\centering
\begin{tabular}{||c|c||c|c||c|c||}
\hline
    $\alpha_0$ & $\beta_0$ & $r\,(N=50)$ & $r\,(N=60)$ &  $n_s\,(N=50)$ & $n_s\,(N=60)$  \\\hline\hline
    $\left(10\sim10^{4}\right)$&$\left(1\sim10^{3}\right)$&$8\times10^{-9}$&$5.5\times10^{-9}$&$0.9600$&$0.9667$\\
    $10^{5}$&$10^{4}$&$7.7\times10^{-9}$&$5.3\times10^{-9}$&$0.9600$&$0.9666$\\
    $10^8$&$10^{7}$&$5.7\times10^{-9}$&$3.7\times10^{-9}$&$0.9530$&$0.9596$\\
    \hline
    \end{tabular}
    \caption{A numerical study of the exact expressions regarding inflationary observables under the assumption of constant values $\xi=10^5$, $\lambda=10^{-4}$ and $\mu=20\,M_P\sim\phi_*$, which together with $\alpha_0$ and $\beta_0$ correctly reproduce the appropriate value of the scalar amplitude $\mathcal{A}_s$. The field excursion $\Delta\phi$ that is presented in the table, is approximated by $\phi_f\approx 10^{-2}M_P$ and $\phi_*\lesssim 20M_P$.}
\end{table}

\begin{figure}[H]
    \centering
    \includegraphics[scale=0.5]{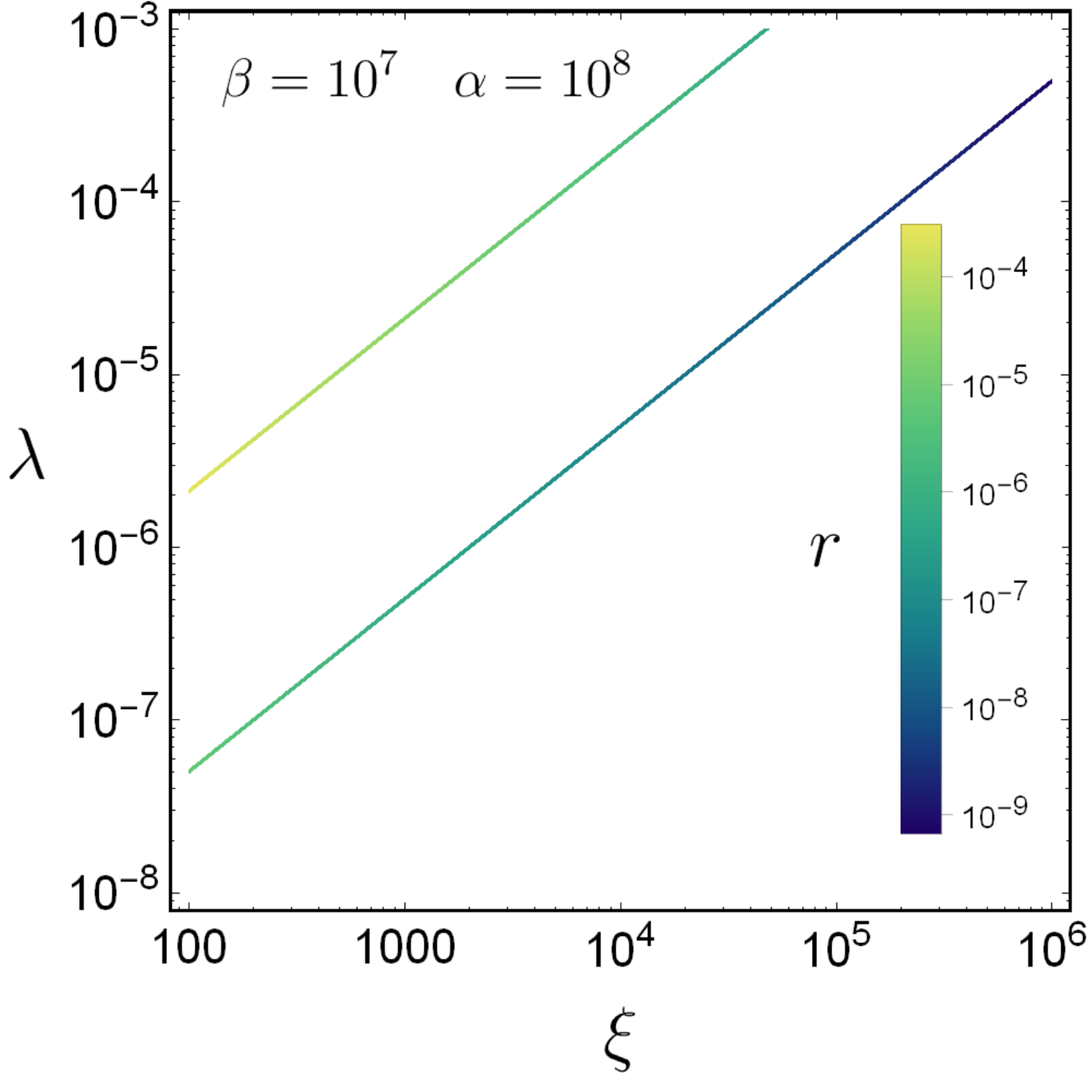}
    \caption{Plot of $\lambda=f(\xi)$. Similar to fig.~\ref{fig1} the values are calculated at $N=50$ $e$-folds and $\mathcal{A}_s\approx2.1\times10^{-9}$. Note the two curves that reproduce the desired value of $\mathcal{A}_s$, attributed to the power interplay between the parameters $\beta_0\lambda$ and $\xi^2$. Depending on their values one of the terms in the expression of the tensor-to-scalar ratio $r$, given by eq.~\eqref{Eq:r}, contributes more, leading to large(r) values of $r$, contrary to fig.~\ref{fig1}.}
    \label{fig2}
\end{figure}

In fig.~\ref{fig2} we notice that increasing the value of the overall coupling $\alpha(\phi)$ leads to an increase in the prediction for the tensor-to-scalar ratio $r$. In fact, the predicted values of $r$ in the region of large $\alpha_0$ and $\beta_0$ are well within the expected accuracy of future experiments $r\sim10^{-4}$~\cite{Matsumura2016, Kogut2011, Sutin2018} and can in principle be separated from other models.

\begin{figure}[H]
    \centering
    \includegraphics[scale=0.4]{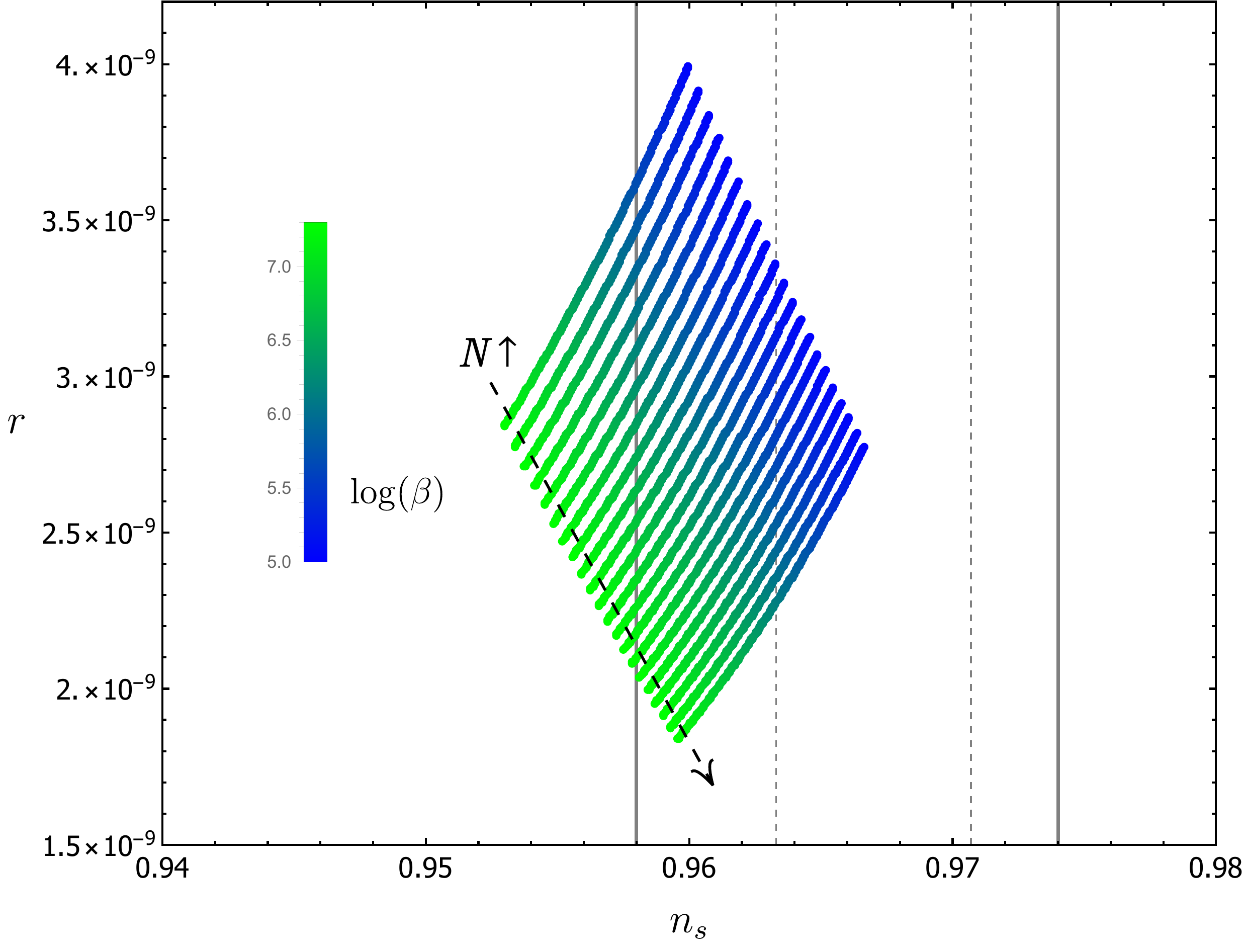}
    \caption{The predictions regarding the inflationary observables in the $r$-$n_s$ plane. The dotted and solid grid lines denote the $1\sigma$ and $2\sigma$ allowed range of $n_s$. We assumed the constant value of $\alpha_0=10^8$, $\xi=2\times10^5$, $\lambda=10^{-4}$ and allowed for varying values of $\beta_0\in\left[10^5,10^7\right]$, as presented in the color grading in the bar of the figure. Note that the number of $e$-foldings are increasing along the direction of the arrow shown above, within the usual range of $N\in\left[50,60\right]$. }
    \label{fig3}
\end{figure}
Larger values of $\beta_0$ decrease the predicted values of $n_s$, made apparent also by eq.~\eqref{Equation:ns}. For values of $\beta_0\sim10^5$ or even less we reach the point where the value of $n_s$ is mainly dependent on the field value of $\phi_*$. Since the coupling constant $\alpha_0$ affects only the value of $r$, in fig.~\ref{fig3} we assumed a constant representative value of $\alpha_0\sim10^8$. Assuming even larger $\alpha_0$ leads to smaller $r$, that is, however, already effectively undetectable.

\begin{figure}[H]
    \centering
    \includegraphics[scale=0.6]{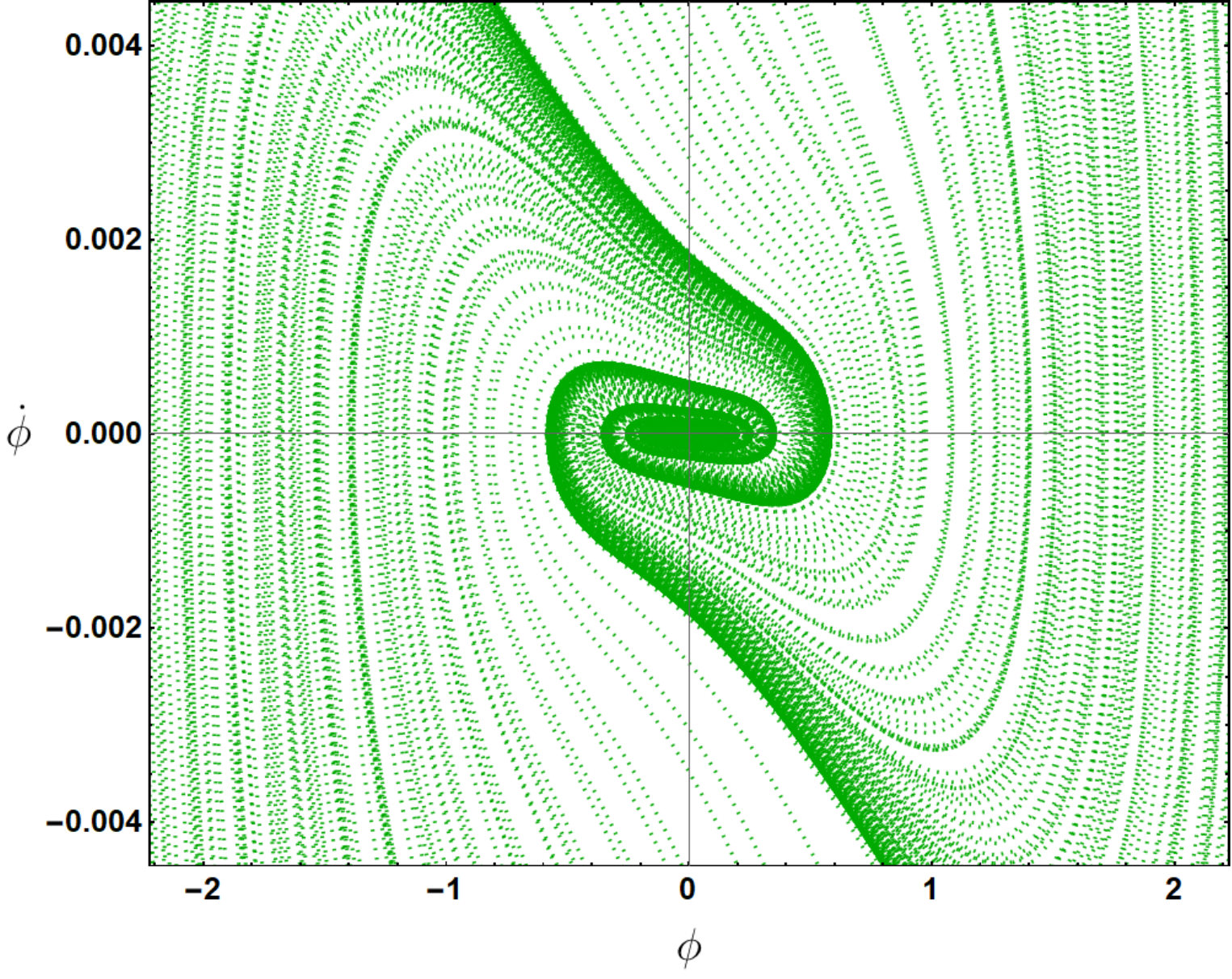}
    \caption{Plot of the phase-space trajectories $\phi$-$\dot{\phi}$ illustrating the attractor point of the potential for constant values of the parameters $\left\{\alpha_0=10^8,\,\beta_0=10^7,\,\xi=2\times10^5,\,\lambda=10^{-4}\right\}$.}
    \label{fig4}
\end{figure}

In the fig.~\ref{fig4} we present the numerical study of the phase-space flow of solution of the generalised Klein-Gordon equation \eqref{Eq:Gen-KG}. For a variety of different initial conditions of the inflaton field the trajectories $\phi$-$\dot{\phi}$ fall into the slow-roll trajectory and end up at the potential minimum, where they are found to oscillate around it. Clearly, the potential has a manifestly attractive behavior and only trajectories that are especially fine-tuned wind up directly into the oscillating phase without any prior amount of slow-roll inflation.

It is worth mentioning that if one disregards the contribution of the higher-order kinetic term $\propto\dot{\phi}^4$ the attractor behaviour of the potential is not spoilt and furthermore the limit of $\beta_0\to0$ correctly reproduces the results of, e.g. refs.~\cite{Antoniadis2019, Tenkanen2020c}, even though not shown in fig.~\eqref{fig3}. It is also the case that small values of $\alpha_0\sim10$, as discussed earlier in the section, allow for a slow-roll solution that serves as an attractor point in the $\phi-\dot{\phi}$ space.

\section{Reheating}\label{section4}

In this section we consider the period of reheating and investigate if the model discussed in the previous section is consistent with reheating, and if possible obtain constraints on the model's parameter space. Without restricting ourselves on any specific reheating paradigm, one is still able to obtain information on the underlying physics through a connection of the reheating and inflationary parameters~\cite{Liddle2003a, Dodelson2003, Dai2014a, Munoz2014, Cook2015a, Gong2015}. The following relation can be expressed as~\cite{Liddle2003a}
\begin{equation}
    \ln{\frac{k}{a_0H_0}}=-N-N_\text{R}-N_\text{eq}+\ln{\frac{a_\text{eq}H_\text{eq}}{a_0H_0}}+\ln{\frac{H_*}{H_\text{eq}}}\,,
\end{equation}
where $k\equiv a_*H_*$ is the comoving Hubble scale. The appearing quantities are defined as
\begin{equation}
    N\equiv\ln{\frac{a_f}{a_*}},\qquad N_{\text{R}}\equiv\ln{\frac{a_{\text{R}}}{a_{f}}},\qquad N_{\text{eq}}\equiv\ln{\frac{a_{\text{eq}}}{a_\text{R}}}\,.
\end{equation}
The amount of $e$-foldings corresponding to the inflationary period is denoted by $N$. Then, $e$-folds denoted by $N_\text{R}$ and $N_\text{eq}$ correspond to the ones from right after the end of inflation ($a_f$) until the end of reheating ($a_\text{R}$), and from the end of reheating to the equilibrium era ($a_\text{eq}$), respectively.

In what follows we assume a constant state parameter $w=p/\rho$ describing each epoch and an abrupt transition between these eras. Assuming conservation of entropy between the transition of reheating to today, and that at the end of inflation the energy density can be approximated\footnote{Corrections to the energy density sourced by the higher-order kinetic term are also associated with deviations of the effective sound speed from unity. In the present article it has been checked numerically that values of ${c_s}^2$, especially near the end of inflation, remain very close to unity (deviations smaller than $10^{-5}$) and we therefore disregard them safely hereafter.} by $\rho_f\approx3U(\phi_f)/2$ one can derive the following expressions for the reheating temperature $T_\text{R}$ and the $e$-folds $N_\text{R}$ as follows~\cite{Cook2015a}
\begin{equation}
    T_\text{R}=\left(\frac{T_\gamma a_0}{k}\right)\left(\frac{43}{11g_\text{R}}\right)^{1/3}H_*e^{-N}e^{-N_\text{R}}\,,
\end{equation}
\begin{equation}\label{Eq:Gen-NR}
    N_\text{R}=\frac{4}{3(1+w_\text{R})}\left\{N+N_\text{R}+\ln{\frac{k}{a_0T_\gamma}}+\ln{\frac{U_f^{1/4}}{H_*}}+\ln{\left[\left(\frac{45}{\pi^2}\right)^{1/4}\left(\frac{11}{3}\right)^{1/3}g_\text{R}^{1/12}\right]}\right\}\,,
\end{equation}
where we defined $U_f\equiv U(\phi_f)$, and $g_\text{R}$ are the relativistic degrees of freedom at the point of reheating.

As expected, an increase in $e$-foldings $N_\text{R}$ implies a decrease in temperature $T_\text{R}$ and vice versa. The actual value of the reheating temperature $T_R$ is bounded from bellow as $T_R> 10\,\text{MeV}$ by BBN, and even temperatures lower than the electroweak scale, assumed for reference here as $T_\text{R}\gtrsim100\,\text{GeV}$, have possible implications with baryogenesis (while not strictly disallowed). An upper bound on the temperature of reheating is usually assumed to be close to the GUT scale of $\sim10^{16}\,\text{GeV}$ in order to avoid the restoration of the GUT symmetry right after inflation.\footnote{In supersymmetric theories the upper bound can be lowered to $\sim10^{9}\,\text{GeV}$~\cite{Moroi1993, Kawasaki1995, Gherghetta1999, Bolz2001}.} Clearly, the known bounds on the reheating temperature are relaxed and cannot provide a better understanding of the reheating era. However, if the state parameter $w_\text{R}$ is known then the actual scenario of reheating can be better inferred. Therefore, we approximate the dynamics of reheating via constant values of $w_\text{R}\in\left\{-\frac{1}{3},1\right\}$, where the value of $-\sfrac{1}{3}$ correspond to the one at the end of inflation and an upper bound of $w_\text{R}\!=\!1$ reflects the complete kinetic domination over the potential.

Regarding the number of $e$-foldings $N_\text{R}$ we can recognise two distinct cases:

\hspace{-1cm}\textbf{Case (a):} Assuming that $w_\text{R}\neq1/3$ we can directly solve eq.~\eqref{Eq:Gen-NR} for $N_\text{R}$ to obtain:
\begin{equation}
    N_\text{R}=\frac{4}{1-3w_\text{R}}\left\{-N-\ln{\frac{U_f^{1/4}}{H_*}}-\ln{\frac{k}{a_0T_\gamma}}-\ln{\left[\left(\frac{45}{\pi^2}\right)^{1/4}\left(\frac{11}{3}\right)^{1/3}g_\text{R}^{1/12}\right]}\right\}\,.
\end{equation}
At reheating, where the temperature $T_\text{R}$ is far greater than the electroweak scale, we have at least the Standard Model degrees of freedom, giving rise to\footnote{Even if degrees of freedom Beyond the Standard Model are included, coming from the far UV-complete region, the number of $e$-folds $N_\text{R}$ and the reheating temperature $T_\text{R}$ remain largely unaffected.} $g_\text{R}\gtrsim\sfrac{427}{4}\approx 100$. At pivot scale $k=0.05\,\text{Mpc}^{-1}$ and $T_\gamma\sim2.7\,\text{K}$ we obtain the simplified version of this:
\begin{equation}\label{Eq:NRnoninst}
    N_\text{R}=\frac{4}{1-3w_\text{R}}\,\left(61.6-\ln{\frac{U_f^{1/4}}{H_*}}-N\right)\,,
\end{equation}
which after some algebra gives rise to the following expression for the temperature 
\begin{equation}\label{Eq:TR-Gen}
    T_\text{R}=\left\{\left(\frac{43}{11g_\text{R}}\right)^{1/3}\left(\frac{T_\gamma a_0}{k}\right)H_*e^{-N}\left(\frac{45}{\pi^2}\,\frac{U_f}{g_\text{R}}\right)^{-\frac{1}{3(1+w_\text{R})}}\right\}^{3(1+w_\text{R})/(3w_\text{R}-1)}\,.
\end{equation}
\textbf{Case (b):} At the special case where $w_\text{R}=1/3$, referred to as instant reheating, we can immediately solve eq.~\eqref{Eq:Gen-NR} to obtain
\begin{equation}\label{Eq:NRinst}
    N=61.6-\ln{\frac{U_f^{1/4}}{H_*}}\,,
\end{equation}
which has effectively become a constrain of the number of $e$-foldings during inflation $N$. This is the assumption of instant reheating, where $N_\text{R}=0$ by default since we instantly assume a radiation dominant era and $T_\text{R}$ takes up its maximum value. 

The focal point is understanding the dynamics of reheating and connecting them to inflationary parameters, which in turn constrains the parameter space discussed in the previous section. Using the definition of the tensor-to-scalar ratio $r\equiv\mathcal{A}_t/\mathcal{A}_s$, with $\mathcal{A}_t=(2H^2)/(\pi^2M_P^2)$, both calculated at the pivot scale $k=a_*H_*$, we obtain:
\begin{equation}
    r_*=\frac{2H_*^2}{\pi^2A_s}\,.
\end{equation}
Then, invoking the slow-roll approximation we can use $r=16\epsilon_V$ in order to finally express $H_*$ in terms of inflationary parameters, i.e.
\begin{equation}
    H_*\approx \pi\sqrt{8A_s\epsilon_*}\,.
\end{equation}
The value of the potential $U(\phi)$ at the end of inflation $\phi=\phi_f$ can be expressed in terms of the slow-roll parameters at the pivot scale.

Let us calculate first the maximum number of $e$-folds assuming instant reheating $w_\text{R}=1/3$. Directly then from eq.~\eqref{Eq:NRinst} we obtain
\begin{equation}\label{Result:Ninst}
    N^\text{inst}_\text{max}\,\approx\, 52\ e\text{-folds}
\end{equation}
specifically for parameter values of $\alpha_0=10$, $\beta_0=1$, $\lambda=10^{-4}$ and $\xi=10^5$ and $\mu\sim20\,M_P$. In fact, the maximum allowed value of $N$ under the assumption of instant reheating is relatively robust to variations of these parameters, with $N^\text{inst}$ tending to values of $51$ $e$-folds for very large $\alpha_0$ and $\beta_0$. 

Allowing for different values of $w_\text{R}$ with the exception of the instant reheating scenario, and using eq.~\eqref{Eq:NRnoninst} together with eq.~\eqref{Eq:TR-Gen} we are able to re-express the reheating temperature as:
\begin{equation}
    T_\text{R}=\left\{\rho_f\left(\frac{30}{\pi^2g_R}\right)\right\}^{1/4}e^{-\frac{3}{4}(1+w_R)N_R}\equiv T_\text{R,max}\,e^{-\frac{3}{4}(1+w_R)N_R}\,,
\end{equation}
where the dependence of the reheating temperature on the $e$-folds $N_\text{R}$ and the state parameter $w_\text{R}$ is made explicit. In eq.~\eqref{Eq:TR-Gen} however, the temperature $T_\text{R}$ is expressed in terms of inflationary parameters, which allows us to parametrise its behaviour with respect to $w_\text{R}$ and varying $e$-folds $N$ in the following figure (fig.~\eqref{fig5}). 

\begin{figure}[H]
    \centering
    \includegraphics[scale=0.35]{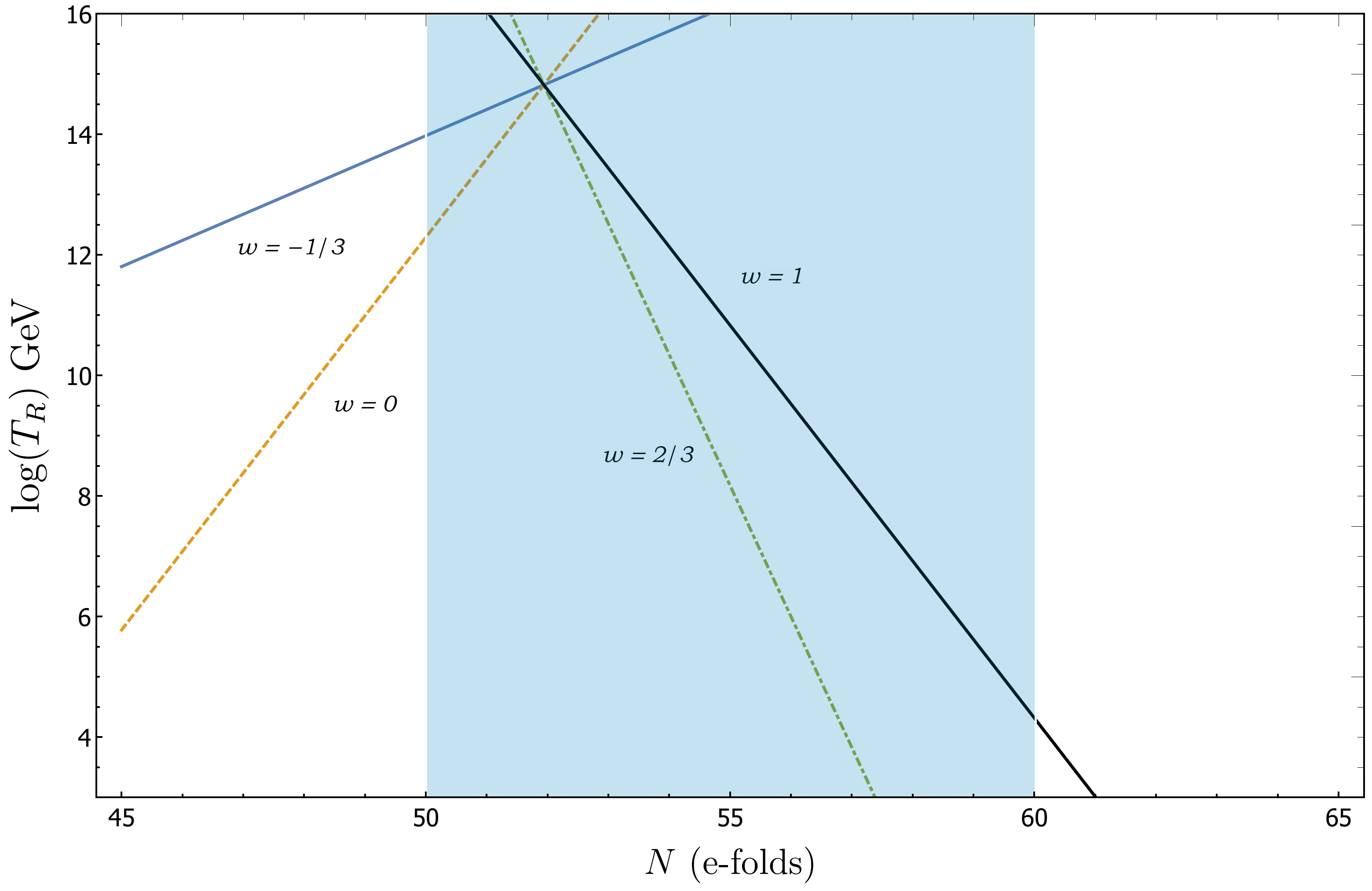}
    \caption{The reheating temperature $T_\text{R}$ plotted against the number of $e$-foldings assumed during the inflationary era, $N$. The colored area denotes the amount of $e$-foldings usually assumed in order to address the issues of early universe cosmology. The blue, dashing-orange, dotted green and black lines represent different values of $w_R\in\left\{\displaystyle{-\frac{1}{3},0,\frac{2}{3},1}\right\}$ respectively.}
    \label{fig5}
\end{figure}
Clearly, all the cases of $w_\text{R}$ considered in fig.~\ref{fig5} converge at the point of instantaneous reheating that would be a vertical line at $N\approx52$ $e$-folds, as discussed in eq.~\eqref{Result:Ninst}. The reheating temperature at the point of convergence is $T_\text{R}\sim 10^{15}\,\text{GeV}$.

\vspace{1cm}
\section{Concluding remarks}\label{section5}

One of the outstanding models with continued success in describing the inflationary phase is the Starobinsky model, $R+R^2$. However, within the Palatini formalism the additional scalar degree of freedom is not propagating, and therefore the model is unable to describe inflation in the conventional way. In this article we investigated the scenario where the Starobinsky model is coupled nonminimally to a real scalar field with a quartic potential. The Starobinsky constant was promoted to also include a dependence on the scalar field, $\alpha\mapsto\alpha(\phi)$, in the form of logarithmic corrections $\log{\left(\phi^2/\mu^2\right)}$.

The resulting action in the Einstein frame contains higher-order kinetic terms of the inflaton field, $(\partial_\mu\phi)^4$, which we ignored during the period of inflation. Different checks were made although, through the complete numerical analysis of the model in order to verify the smallness of their contribution. The potential also exhibits a plateau, attributed to the $R^2$ term, that can provide a suitable amount of inflation. In contrast to the regular Palatini-$R^2$ models~\cite{Antoniadis2018, Antoniadis2019, Enckell2019, Gialamas2020, LloydStubbs2020} the plateau in this case is violated logarithmically at scales where $\phi>\mu$. The inflationary predictions of the model are in agreement with observational data for a certain region of the free parameter space, which also maintained the attractor point of the potential. In that same allowed region, for larger values of the constants $\alpha_0$ and $\beta_0$ we found that the field dependence of the coupling $\alpha(\phi)$ influences the inflationary observables, most importantly $n_s$, as seen in fig.~\ref{fig3}. It is safe to expect that by introducing a similar coupling $\alpha(\phi)$ in other models with predictions lying outside the $+2\sigma$ region of $n_s$, one can potentially obtain values that are in agreement with observational bounds. Notably the predicted values of the tensor-to-scalar ratio range from tiny (a feature of the Palatini-$R^2$ models) to rather large $\sim10^{-4}$ that can in principle come in contact with future experiments with expected precision of $10^{-3}$ or $10^{-4}$. 

Following the end of inflation we explored the possibility of the model undergoing the process of reheating. Through a parametrisation of the reheating parameters in terms of the inflationary ones, based on ref.~\cite{Liddle2003a}, we showed that the model can indeed support a reheating scenario alongside a successful inflationary era. In addition, under the assumption of instantaneous reheating the upper bound of $N\approx52$ $e$-folds during inflation is obtained, while for varying values of the state parameter $w_\text{R}\in\left\{-\sfrac{1}{3},1\right\}$ we found that the maximum reheating temperature is $T_\text{R}\sim10^{15}\,\text{GeV}$, with lower values also achievable within a reasonable amount of $e$-foldings $N$. 

\vspace{0.25cm}
\subsection*{Acknowledgements}
 The research of A.L. is co-financed by Greece and the European Union (European Social Fund - ESF) through the Operational Programme “Human Resources Development, Education and Lifelong Learning” in the context of the project “Strengthening Human Resources Research Potential via Doctorate Research” (MIS-5000432), implemented by the State Scholarships Foundation (IKY).
\vspace{1cm}

\bibliography{refs}{}

\providecommand{\href}[2]{#2}\begingroup\raggedright\begin{thebibliography}{10}

\bibitem{Guth1981}
A.~H. Guth, {\it {The Inflationary Universe: A Possible Solution to the Horizon
  and Flatness Problems}},  {\em Phys. Rev.} {\bf D23} (1981) 347--356.

\bibitem{Linde1982}
A.~D. Linde, {\it {A New Inflationary Universe Scenario: A Possible Solution of
  the Horizon, Flatness, Homogeneity, Isotropy and Primordial Monopole
  Problems}},  {\em Phys. Lett.} {\bf 108B} (1982) 389--393.

\bibitem{Albrecht1982}
A.~Albrecht and P.~J. Steinhardt, {\it {Cosmology for Grand Unified Theories
  with Radiatively Induced Symmetry Breaking}},  {\em Phys. Rev. Lett.} {\bf
  48} (1982) 1220--1223.

\bibitem{Sato1981}
K.~Sato, {\it {First Order Phase Transition of a Vacuum and Expansion of the
  Universe}},  {\em Mon. Not. Roy. Astron. Soc.} {\bf 195} (1981) 467--479.

\bibitem{Linde1983a}
A.~D. Linde, {\it {Chaotic Inflation}},  {\em Phys. Lett.} {\bf 129B} (1983)
  177--181.

\bibitem{Starobinsky1979}
A.~A. Starobinsky, {\it {Spectrum of relict gravitational radiation and the
  early state of the universe}},  {\em JETP Lett.} {\bf 30} (1979) 682--685.
  [Pisma Zh. Eksp. Teor. Fiz.30,719(1979); ,767(1979)].

\bibitem{Mukhanov1981}
V.~F. Mukhanov and G.~V. Chibisov, {\it {Quantum Fluctuations and a Nonsingular
  Universe}},  {\em JETP Lett.} {\bf 33} (1981) 532--535. [Pisma Zh. Eksp.
  Teor. Fiz.33,549(1981)].

\bibitem{Hawking1982}
S.~W. Hawking, {\it {The Development of Irregularities in a Single Bubble
  Inflationary Universe}},  {\em Phys. Lett.} {\bf 115B} (1982) 295.

\bibitem{Hawking1983}
S.~W. Hawking and I.~G. Moss, {\it {Fluctuations in the Inflationary
  Universe}},  {\em Nucl. Phys.} {\bf B224} (1983) 180.

\bibitem{Starobinsky1982}
A.~A. Starobinsky, {\it {Dynamics of Phase Transition in the New Inflationary
  Universe Scenario and Generation of Perturbations}},  {\em Phys. Lett.} {\bf
  117B} (1982) 175--178.

\bibitem{Guth1982}
A.~H. Guth and S.~Y. Pi, {\it {Fluctuations in the New Inflationary Universe}},
   {\em Phys. Rev. Lett.} {\bf 49} (1982) 1110--1113.

\bibitem{Starobinsky1980}
A.~A. Starobinsky, {\it {A New Type of Isotropic Cosmological Models Without
  Singularity}},  {\em Phys. Lett.} {\bf 91B} (1980) 99--102.

\bibitem{Bezrukov2008}
F.~L. Bezrukov and M.~Shaposhnikov, {\it {The Standard Model Higgs boson as the
  inflaton}},  {\em Phys. Lett.} {\bf B659} (2008) 703--706,
  [\href{http://arxiv.org/abs/0710.3755}{{\tt arXiv:0710.3755}}].

\bibitem{DeSimone2009}
A.~De~Simone, M.~P. Hertzberg, and F.~Wilczek, {\it {Running Inflation in the
  Standard Model}},  {\em Phys. Lett.} {\bf B678} (2009) 1--8,
  [\href{http://arxiv.org/abs/0812.4946}{{\tt arXiv:0812.4946}}].

\bibitem{Bezrukov2009a}
F.~L. Bezrukov, A.~Magnin, and M.~Shaposhnikov, {\it {Standard Model Higgs
  boson mass from inflation}},  {\em Phys. Lett.} {\bf B675} (2009) 88--92,
  [\href{http://arxiv.org/abs/0812.4950}{{\tt arXiv:0812.4950}}].

\bibitem{Akrami2018}
{\bf Planck} Collaboration, Y.~Akrami et~al., {\it {Planck 2018 results. X.
  Constraints on inflation}},  \href{http://arxiv.org/abs/1807.06211}{{\tt
  arXiv:1807.06211}}.

\bibitem{Ade2018}
{\bf BICEP2, Keck Array} Collaboration, P.~A.~R. Ade et~al., {\it {BICEP2 /
  Keck Array x: Constraints on Primordial Gravitational Waves using Planck,
  WMAP, and New BICEP2/Keck Observations through the 2015 Season}},  {\em Phys.
  Rev. Lett.} {\bf 121} (2018) 221301,
  [\href{http://arxiv.org/abs/1810.05216}{{\tt arXiv:1810.05216}}].

\bibitem{Abbott1982}
L.~F. Abbott, E.~Farhi, and M.~B. Wise, {\it {Particle Production in the New
  Inflationary Cosmology}},  {\em Phys. Lett. B} {\bf 117} (1982) 29.

\bibitem{Dolgov1982}
A.~D. Dolgov and A.~D. Linde, {\it {Baryon Asymmetry in Inflationary
  Universe}},  {\em Phys. Lett. B} {\bf 116} (1982) 329.

\bibitem{Albrecht1982b}
A.~Albrecht, P.~J. Steinhardt, M.~S. Turner, and F.~Wilczek, {\it {Reheating an
  Inflationary Universe}},  {\em Phys. Rev. Lett.} {\bf 48} (1982) 1437.

\bibitem{Felder1999}
G.~N. Felder, L.~Kofman, and A.~D. Linde, {\it {Instant preheating}},  {\em
  Phys. Rev. D} {\bf 59} (1999) 123523,
  [\href{http://arxiv.org/abs/hep-ph/9812289}{{\tt hep-ph/9812289}}].

\bibitem{Kofman1994}
L.~Kofman, A.~D. Linde, and A.~A. Starobinsky, {\it {Reheating after
  inflation}},  {\em Phys. Rev. Lett.} {\bf 73} (1994) 3195--3198,
  [\href{http://arxiv.org/abs/hep-th/9405187}{{\tt hep-th/9405187}}].

\bibitem{Traschen1990}
J.~H. Traschen and R.~H. Brandenberger, {\it {Particle Production During
  Out-of-equilibrium Phase Transitions}},  {\em Phys. Rev. D} {\bf 42} (1990)
  2491--2504.

\bibitem{Kofman1997}
L.~Kofman, A.~D. Linde, and A.~A. Starobinsky, {\it {Towards the theory of
  reheating after inflation}},  {\em Phys. Rev. D} {\bf 56} (1997) 3258--3295,
  [\href{http://arxiv.org/abs/hep-ph/9704452}{{\tt hep-ph/9704452}}].

\bibitem{Greene1997}
B.~R. Greene, T.~Prokopec, and T.~G. Roos, {\it {Inflaton decay and heavy
  particle production with negative coupling}},  {\em Phys. Rev. D} {\bf 56}
  (1997) 6484--6507, [\href{http://arxiv.org/abs/hep-ph/9705357}{{\tt
  hep-ph/9705357}}].

\bibitem{Felder2001}
G.~N. Felder, J.~Garcia-Bellido, P.~B. Greene, L.~Kofman, A.~D. Linde, and
  I.~Tkachev, {\it {Dynamics of symmetry breaking and tachyonic preheating}},
  {\em Phys. Rev. Lett.} {\bf 87} (2001) 011601,
  [\href{http://arxiv.org/abs/hep-ph/0012142}{{\tt hep-ph/0012142}}].

\bibitem{Felder2001a}
G.~N. Felder, L.~Kofman, and A.~D. Linde, {\it {Tachyonic instability and
  dynamics of spontaneous symmetry breaking}},  {\em Phys. Rev. D} {\bf 64}
  (2001) 123517, [\href{http://arxiv.org/abs/hep-th/0106179}{{\tt
  hep-th/0106179}}].

\bibitem{Shuhmaher2006}
N.~Shuhmaher and R.~Brandenberger, {\it {Non-perturbative instabilities as a
  solution of the cosmological moduli problem}},  {\em Phys. Rev. D} {\bf 73}
  (2006) 043519, [\href{http://arxiv.org/abs/hep-th/0507103}{{\tt
  hep-th/0507103}}].

\bibitem{Dufaux2006}
J.~F. Dufaux, G.~N. Felder, L.~Kofman, M.~Peloso, and D.~Podolsky, {\it
  {Preheating with trilinear interactions: Tachyonic resonance}},  {\em JCAP}
  {\bf 07} (2006) 006, [\href{http://arxiv.org/abs/hep-ph/0602144}{{\tt
  hep-ph/0602144}}].

\bibitem{Dodelson2003}
S.~Dodelson and L.~Hui, {\it {A Horizon ratio bound for inflationary
  fluctuations}},  {\em Phys. Rev. Lett.} {\bf 91} (2003) 131301,
  [\href{http://arxiv.org/abs/astro-ph/0305113}{{\tt astro-ph/0305113}}].

\bibitem{Liddle2003a}
A.~R. Liddle and S.~M. Leach, {\it {How long before the end of inflation were
  observable perturbations produced?}},  {\em Phys. Rev. D} {\bf 68} (2003)
  103503, [\href{http://arxiv.org/abs/astro-ph/0305263}{{\tt
  astro-ph/0305263}}].

\bibitem{Dai2014a}
L.~Dai, M.~Kamionkowski, and J.~Wang, {\it {Reheating constraints to
  inflationary models}},  {\em Phys. Rev. Lett.} {\bf 113} (2014) 041302,
  [\href{http://arxiv.org/abs/1404.6704}{{\tt arXiv:1404.6704}}].

\bibitem{Munoz2014}
J.~B. Munoz and M.~Kamionkowski, {\it {Equation-of-State Parameter for
  Reheating}},  {\em Phys. Rev. D 91, 043521 (2015)} (Dec., 2014)
  [\href{http://arxiv.org/abs/1412.0656}{{\tt arXiv:1412.0656}}].

\bibitem{Gong2015}
J.-O. Gong, S.~Pi, and G.~Leung, {\it {Probing reheating with primordial
  spectrum}},  {\em JCAP} {\bf 05} (2015) 027,
  [\href{http://arxiv.org/abs/1501.03604}{{\tt arXiv:1501.03604}}].

\bibitem{Cook2015a}
J.~L. Cook, E.~Dimastrogiovanni, D.~A. Easson, and L.~M. Krauss, {\it
  {Reheating predictions in single field inflation}},  {\em JCAP} {\bf 04}
  (2015) 047, [\href{http://arxiv.org/abs/1502.04673}{{\tt arXiv:1502.04673}}].

\bibitem{Palatini:1919}
A.~Palatini, {\it Deduzione invariantiva delle equazioni gravitazionali dal
  principio di hamilton},  {\em Rendiconti del Circolo Matematico di Palermo
  (1884-1940)} {\bf 43} (Dec, 1919) 203--212.

\bibitem{Bauer2008}
F.~Bauer and D.~A. Demir, {\it {Inflation with Non-Minimal Coupling: Metric
  versus Palatini Formulations}},  {\em Phys. Lett.} {\bf B665} (2008)
  222--226, [\href{http://arxiv.org/abs/0803.2664}{{\tt arXiv:0803.2664}}].

\bibitem{Bauer2011b}
F.~Bauer, {\it {Filtering out the cosmological constant in the Palatini
  formalism of modified gravity}},  {\em Gen. Rel. Grav.} {\bf 43} (2011)
  1733--1757, [\href{http://arxiv.org/abs/1007.2546}{{\tt arXiv:1007.2546}}].

\bibitem{Tamanini2011}
N.~Tamanini and C.~R. Contaldi, {\it {Inflationary Perturbations in Palatini
  Generalised Gravity}},  {\em Phys. Rev.} {\bf D83} (2011) 044018,
  [\href{http://arxiv.org/abs/1010.0689}{{\tt arXiv:1010.0689}}].

\bibitem{Bauer2011}
F.~Bauer and D.~A. Demir, {\it {Higgs-Palatini Inflation and Unitarity}},  {\em
  Phys. Lett.} {\bf B698} (2011) 425--429,
  [\href{http://arxiv.org/abs/1012.2900}{{\tt arXiv:1012.2900}}].

\bibitem{Borowiec2012}
A.~Borowiec, M.~Kamionka, A.~Kurek, and M.~Szydlowski, {\it {Cosmic
  acceleration from modified gravity with Palatini formalism}},  {\em JCAP}
  {\bf 1202} (2012) 027, [\href{http://arxiv.org/abs/1109.3420}{{\tt
  arXiv:1109.3420}}].

\bibitem{Stachowski2017}
A.~Stachowski, M.~Szydłowski, and A.~Borowiec, {\it {Starobinsky cosmological
  model in Palatini formalism}},  {\em Eur. Phys. J.} {\bf C77} (2017), no.~6
  406, [\href{http://arxiv.org/abs/1608.03196}{{\tt arXiv:1608.03196}}].

\bibitem{Rasanen2017}
S.~Rasanen and P.~Wahlman, {\it {Higgs inflation with loop corrections in the
  Palatini formulation}},  {\em JCAP} {\bf 1711} (2017), no.~11 047,
  [\href{http://arxiv.org/abs/1709.07853}{{\tt arXiv:1709.07853}}].

\bibitem{Tenkanen2017}
T.~Tenkanen, {\it {Resurrecting Quadratic Inflation with a non-minimal coupling
  to gravity}},  {\em JCAP} {\bf 1712} (2017), no.~12 001,
  [\href{http://arxiv.org/abs/1710.02758}{{\tt arXiv:1710.02758}}].

\bibitem{Racioppi2017}
A.~Racioppi, {\it {Coleman-Weinberg linear inflation: metric vs. Palatini
  formulation}},  {\em JCAP} {\bf 1712} (2017), no.~12 041,
  [\href{http://arxiv.org/abs/1710.04853}{{\tt arXiv:1710.04853}}].

\bibitem{Markkanen2018}
T.~Markkanen, T.~Tenkanen, V.~Vaskonen, and H.~Veermäe, {\it {Quantum
  corrections to quartic inflation with a non-minimal coupling: metric vs.
  Palatini}},  {\em JCAP} {\bf 1803} (2018), no.~03 029,
  [\href{http://arxiv.org/abs/1712.04874}{{\tt arXiv:1712.04874}}].

\bibitem{Jaerv2018}
L.~J\"{a}rv, A.~Racioppi, and T.~Tenkanen, {\it {Palatini side of inflationary
  attractors}},  {\em Phys. Rev.} {\bf D97} (2018), no.~8 083513,
  [\href{http://arxiv.org/abs/1712.08471}{{\tt arXiv:1712.08471}}].

\bibitem{Fu2017}
C.~Fu, P.~Wu, and H.~Yu, {\it {Inflationary dynamics and preheating of the
  nonminimally coupled inflaton field in the metric and Palatini formalisms}},
  {\em Phys. Rev.} {\bf D96} (2017), no.~10 103542,
  [\href{http://arxiv.org/abs/1801.04089}{{\tt arXiv:1801.04089}}].

\bibitem{Enckell2018}
V.-M. Enckell, K.~Enqvist, S.~Rasanen, and E.~Tomberg, {\it {Higgs inflation at
  the hilltop}},  {\em JCAP} {\bf 1806} (2018), no.~06 005,
  [\href{http://arxiv.org/abs/1802.09299}{{\tt arXiv:1802.09299}}].

\bibitem{Kozak2019}
A.~Kozak and A.~Borowiec, {\it Palatini frames in scalar–tensor theories of
  gravity},  {\em Eur. Phys. J. C} {\bf 79} (2019), no.~4 335,
  [\href{http://arxiv.org/abs/1808.05598}{{\tt arXiv:1808.05598}}].

\bibitem{Enckell2019}
V.-M. Enckell, K.~Enqvist, S.~Rasanen, and L.-P. Wahlman, {\it {Inflation with
  $R^2$ term in the Palatini formalism}},  {\em JCAP} {\bf 02} (2019) 022,
  [\href{http://arxiv.org/abs/1810.05536}{{\tt arXiv:1810.05536}}].

\bibitem{Antoniadis2018}
I.~Antoniadis, A.~Karam, A.~Lykkas, and K.~Tamvakis, {\it {Palatini inflation
  in models with an $R^2$ term}},  {\em JCAP} {\bf 1811} (2018), no.~11 028,
  [\href{http://arxiv.org/abs/1810.10418}{{\tt arXiv:1810.10418}}].

\bibitem{Rasanen2019a}
S.~Rasanen and E.~Tomberg, {\it {Planck scale black hole dark matter from Higgs
  inflation}},  {\em JCAP} {\bf 01} (2019) 038,
  [\href{http://arxiv.org/abs/1810.12608}{{\tt arXiv:1810.12608}}].

\bibitem{Rasanen2019}
S.~Rasanen, {\it {Higgs inflation in the Palatini formulation with kinetic
  terms for the metric}},  {\em Open J. Astrophys.} {\bf 2} (2019), no.~1 1,
  [\href{http://arxiv.org/abs/1811.09514}{{\tt arXiv:1811.09514}}].

\bibitem{Almeida2019}
J.~P.~B. Almeida, N.~Bernal, J.~Rubio, and T.~Tenkanen, {\it {Hidden Inflaton
  Dark Matter}},  {\em JCAP} {\bf 03} (2019) 012,
  [\href{http://arxiv.org/abs/1811.09640}{{\tt arXiv:1811.09640}}].

\bibitem{Antoniadis2019}
I.~Antoniadis, A.~Karam, A.~Lykkas, T.~Pappas, and K.~Tamvakis, {\it {Rescuing
  Quartic and Natural Inflation in the Palatini Formalism}},  {\em JCAP} {\bf
  03} (2019) 005, [\href{http://arxiv.org/abs/1812.00847}{{\tt
  arXiv:1812.00847}}].

\bibitem{Jinno2019}
R.~Jinno, K.~Kaneta, K.-y. Oda, and S.~C. Park, {\it {Hillclimbing inflation in
  metric and Palatini formulations}},  {\em Phys. Lett.} {\bf B791} (2019)
  396--402, [\href{http://arxiv.org/abs/1812.11077}{{\tt arXiv:1812.11077}}].

\bibitem{Tenkanen2019a}
T.~Tenkanen, {\it {Minimal Higgs inflation with an $R^2$ term in Palatini
  gravity}},  {\em Phys. Rev. D} {\bf 99} (2019), no.~6 063528,
  [\href{http://arxiv.org/abs/1901.01794}{{\tt arXiv:1901.01794}}].

\bibitem{Edery2019}
A.~Edery and Y.~Nakayama, {\it {Palatini formulation of pure $R^2$ gravity
  yields Einstein gravity with no massless scalar}},  {\em Phys. Rev. D} {\bf
  99} (2019), no.~12 124018, [\href{http://arxiv.org/abs/1902.07876}{{\tt
  arXiv:1902.07876}}].

\bibitem{Rubio2019}
J.~Rubio and E.~S. Tomberg, {\it {Preheating in Palatini Higgs inflation}},
  {\em JCAP} {\bf 1904} (2019), no.~04 021,
  [\href{http://arxiv.org/abs/1902.10148}{{\tt arXiv:1902.10148}}].

\bibitem{Tenkanen2019b}
T.~Tenkanen and L.~Visinelli, {\it {Axion dark matter from Higgs inflation with
  an intermediate $H_*$}},  {\em JCAP} {\bf 08} (2019) 033,
  [\href{http://arxiv.org/abs/1906.11837}{{\tt arXiv:1906.11837}}].

\bibitem{Bostan2020}
N.~Bostan, {\it {Non-minimally coupled quartic inflation with Coleman-Weinberg
  one-loop corrections in the Palatini formulation}},  {\em Phys. Lett. B} {\bf
  811} (2020) 135954, [\href{http://arxiv.org/abs/1907.13235}{{\tt
  arXiv:1907.13235}}].

\bibitem{Tenkanen2020d}
T.~Tenkanen, {\it {Trans-Planckian censorship, inflation, and dark matter}},
  {\em Phys. Rev. D} {\bf 101} (2020), no.~6 063517,
  [\href{http://arxiv.org/abs/1910.00521}{{\tt arXiv:1910.00521}}].

\bibitem{Gialamas2020}
I.~D. Gialamas and A.~B. Lahanas, {\it {Reheating in $R^2$ Palatini
  inflationary models}},  {\em Phys. Rev. D} {\bf 101} (2020), no.~8 084007,
  [\href{http://arxiv.org/abs/1911.11513}{{\tt arXiv:1911.11513}}].

\bibitem{Racioppi2020}
A.~Racioppi, {\it Non-minimal (self-)running inflation: Metric vs. palatini
  formulation},  {\em JHEP} {\bf 21} (2020) 011,
  [\href{http://arxiv.org/abs/1912.10038}{{\tt arXiv:1912.10038}}].

\bibitem{Shaposhnikov2021}
M.~Shaposhnikov, A.~Shkerin, and S.~Zell, {\it {Standard Model Meets Gravity:
  Electroweak Symmetry Breaking and Inflation}},  {\em Phys. Rev. D} {\bf 103}
  (2021), no.~3 033006, [\href{http://arxiv.org/abs/2001.09088}{{\tt
  arXiv:2001.09088}}].

\bibitem{Tenkanen2020b}
T.~Tenkanen, {\it {Tracing the high energy theory of gravity: an introduction
  to Palatini inflation}},  {\em Gen. Rel. Grav.} {\bf 52} (2020), no.~4 33,
  [\href{http://arxiv.org/abs/2001.10135}{{\tt arXiv:2001.10135}}].

\bibitem{Tenkanen2020c}
T.~Tenkanen and E.~Tomberg, {\it {Initial conditions for plateau inflation: a
  case study}},  {\em JCAP} {\bf 04} (2020) 050,
  [\href{http://arxiv.org/abs/2002.02420}{{\tt arXiv:2002.02420}}].

\bibitem{Shaposhnikov2020b}
M.~Shaposhnikov, A.~Shkerin, and S.~Zell, {\it {Quantum Effects in Palatini
  Higgs Inflation}},  {\em JCAP} {\bf 07} (2020) 064,
  [\href{http://arxiv.org/abs/2002.07105}{{\tt arXiv:2002.07105}}].

\bibitem{LloydStubbs2020}
A.~Lloyd-Stubbs and J.~McDonald, {\it {Sub-Planckian $\phi^2$ inflation in the
  Palatini formulation of gravity with an $R^2$ term}},  {\em Phys. Rev. D}
  {\bf 101} (2020), no.~12 123515, [\href{http://arxiv.org/abs/2002.08324}{{\tt
  arXiv:2002.08324}}].

\bibitem{Antoniadis2020}
I.~Antoniadis, A.~Lykkas, and K.~Tamvakis, {\it {Constant-roll in the
  Palatini-$R^2$ models}},  {\em JCAP} {\bf 04} (2020), no.~04 033,
  [\href{http://arxiv.org/abs/2002.12681}{{\tt arXiv:2002.12681}}].

\bibitem{Ghilencea2020a}
D.~M. Ghilencea, {\it {Palatini quadratic gravity: spontaneous breaking of
  gauged scale symmetry and inflation}},  {\em Eur. Phys. J. C} {\bf 80} (4,
  2020) 1147, [\href{http://arxiv.org/abs/2003.08516}{{\tt arXiv:2003.08516}}].

\bibitem{Das2020}
N.~Das and S.~Panda, {\it {Inflation and Reheating in $f(R,h)$ theory
  formulated in the Palatini formalism}},
  \href{http://arxiv.org/abs/2005.14054}{{\tt arXiv:2005.14054}}.

\bibitem{Jaerv2020}
L.~J\"arv, A.~Karam, A.~Kozak, A.~Lykkas, A.~Racioppi, and M.~Saal, {\it
  {Equivalence of inflationary models between the metric and Palatini
  formulation of scalar-tensor theories}},  {\em Phys. Rev. D} {\bf 102}
  (2020), no.~4 044029, [\href{http://arxiv.org/abs/2005.14571}{{\tt
  arXiv:2005.14571}}].

\bibitem{Gialamas2020b}
I.~D. Gialamas, A.~Karam, and A.~Racioppi, {\it {Dynamically induced Planck
  scale and inflation in the Palatini formulation}},  {\em JCAP} {\bf 11}
  (2020) 014, [\href{http://arxiv.org/abs/2006.09124}{{\tt arXiv:2006.09124}}].

\bibitem{Karam2020}
A.~Karam, M.~Raidal, and E.~Tomberg, {\it {Gravitational dark matter production
  in Palatini preheating}},  \href{http://arxiv.org/abs/2007.03484}{{\tt
  arXiv:2007.03484}}.

\bibitem{McDonald2020}
J.~McDonald, {\it {Does Palatini Higgs Inflation Conserve Unitarity?}},
  \href{http://arxiv.org/abs/2007.04111}{{\tt arXiv:2007.04111}}.

\bibitem{Laangvik2020}
M.~L\r{a}ngvik, J.-M. Ojanper\"a, S.~Raatikainen, and S.~Rasanen, {\it {Higgs
  inflation with the Holst and the Nieh-Yan term}},
  \href{http://arxiv.org/abs/2007.12595}{{\tt arXiv:2007.12595}}.

\bibitem{Ghilencea2020}
D.~M. Ghilencea, {\it {Gauging scale symmetry and inflation: Weyl versus
  Palatini gravity}},  \href{http://arxiv.org/abs/2007.14733}{{\tt
  arXiv:2007.14733}}.

\bibitem{Shaposhnikov2021a}
M.~Shaposhnikov, A.~Shkerin, I.~Timiryasov, and S.~Zell, {\it {Higgs inflation
  in Einstein-Cartan gravity}},  {\em JCAP} {\bf 02} (2021) 008,
  [\href{http://arxiv.org/abs/2007.14978}{{\tt arXiv:2007.14978}}].

\bibitem{Shaposhnikov2020c}
M.~Shaposhnikov, A.~Shkerin, I.~Timiryasov, and S.~Zell, {\it {Einstein-Cartan
  gravity, matter, and scale-invariant generalization}},  {\em JHEP} {\bf 10}
  (2020) 177, [\href{http://arxiv.org/abs/2007.16158}{{\tt arXiv:2007.16158}}].

\bibitem{Gialamas2020a}
I.~D. Gialamas, A.~Karam, A.~Lykkas, and T.~D. Pappas, {\it {Palatini-Higgs
  inflation with nonminimal derivative coupling}},  {\em Phys. Rev. D} {\bf
  102} (2020), no.~6 063522, [\href{http://arxiv.org/abs/2008.06371}{{\tt
  arXiv:2008.06371}}].

\bibitem{Verner2020}
S.~Verner, {\it {Quintessential Inflation in Palatini Gravity}},
  \href{http://arxiv.org/abs/2010.11201}{{\tt arXiv:2010.11201}}.

\bibitem{Bekov2020}
S.~Bekov, K.~Myrzakulov, R.~Myrzakulov, and D.~S.-C. G\'omez, {\it {General
  slow-roll inflation in $f(R)$ gravity under the Palatini approach}},  {\em
  Symmetry} {\bf 12} (2020), no.~12 1958,
  [\href{http://arxiv.org/abs/2010.12360}{{\tt arXiv:2010.12360}}].

\bibitem{Enckell2020}
V.-M. Enckell, S.~Nurmi, S.~Rasanen, and E.~Tomberg, {\it {Critical point Higgs
  inflation in the Palatini formulation}},
  \href{http://arxiv.org/abs/2012.03660}{{\tt arXiv:2012.03660}}.

\bibitem{Dimopoulos2021}
K.~Dimopoulos and S.~S\'anchez~L\'opez, {\it {Quintessential inflation in
  Palatini $f(R)$ gravity}},  {\em Phys. Rev. D} {\bf 103} (2021), no.~4
  043533, [\href{http://arxiv.org/abs/2012.06831}{{\tt arXiv:2012.06831}}].

\bibitem{Karam2021a}
A.~Karam, E.~Tomberg, and H.~Veerm\"ae, {\it {Tachyonic Preheating in Palatini
  $R^2$ Inflation}},  \href{http://arxiv.org/abs/2102.02712}{{\tt
  arXiv:2102.02712}}.

\bibitem{Karam2021}
A.~Karam, S.~Karamitsos, and M.~Saal, {\it {$\beta$-function reconstruction of
  Palatini inflationary attractors}},
  \href{http://arxiv.org/abs/2103.01182}{{\tt arXiv:2103.01182}}.

\bibitem{Parker2009}
L.~E. Parker and D.~Toms, {\em {Quantum Field Theory in Curved Spacetime:
  Quantized Field and Gravity}}.
\newblock Cambridge Monographs on Mathematical Physics. Cambridge University
  Press, 8, 2009.

\bibitem{Lyth1999}
D.~H. Lyth and A.~Riotto, {\it {Particle physics models of inflation and the
  cosmological density perturbation}},  {\em Physics Reports} {\bf 314} (1999),
  no.~1 1--146.

\bibitem{Gundhi2021}
A.~Gundhi, S.~V. Ketov, and C.~F. Steinwachs, {\it {Primordial black hole dark
  matter in dilaton-extended two-field Starobinsky inflation}},  {\em Phys.
  Rev. D} {\bf 103} (2021), no.~8 083518,
  [\href{http://arxiv.org/abs/2011.05999}{{\tt arXiv:2011.05999}}].

\bibitem{Gundhi2021a}
A.~Gundhi and C.~F. Steinwachs, {\it {Scalaron–Higgs inflation reloaded:
  Higgs-dependent scalaron mass and primordial black hole dark matter}},  {\em
  Eur. Phys. J. C} {\bf 81} (2021), no.~5 460,
  [\href{http://arxiv.org/abs/2011.09485}{{\tt arXiv:2011.09485}}].

\bibitem{Matsumura2016}
T.~Matsumura et~al., {\it {LiteBIRD: Mission Overview and Focal Plane Layout}},
   {\em J. Low Temp. Phys.} {\bf 184} (2016), no.~3-4 824--831.

\bibitem{Kogut2011}
A.~Kogut et~al., {\it {The Primordial Inflation Explorer (PIXIE): A Nulling
  Polarimeter for Cosmic Microwave Background Observations}},  {\em JCAP} {\bf
  07} (2011) 025, [\href{http://arxiv.org/abs/1105.2044}{{\tt
  arXiv:1105.2044}}].

\bibitem{Sutin2018}
B.~M. Sutin et~al., {\it {PICO - the probe of inflation and cosmic origins}},
  {\em Proc. SPIE Int. Soc. Opt. Eng.} {\bf 10698} (2018) 106984F,
  [\href{http://arxiv.org/abs/1808.01368}{{\tt arXiv:1808.01368}}].

\bibitem{Moroi1993}
T.~Moroi, H.~Murayama, and M.~Yamaguchi, {\it {Cosmological constraints on the
  light stable gravitino}},  {\em Phys. Lett. B} {\bf 303} (1993) 289--294.

\bibitem{Kawasaki1995}
M.~Kawasaki and T.~Moroi, {\it {Gravitino production in the inflationary
  universe and the effects on big bang nucleosynthesis}},  {\em Prog. Theor.
  Phys.} {\bf 93} (1995) 879--900,
  [\href{http://arxiv.org/abs/hep-ph/9403364}{{\tt hep-ph/9403364}}].

\bibitem{Gherghetta1999}
T.~Gherghetta, G.~F. Giudice, and A.~Riotto, {\it {Nucleosynthesis bounds in
  gauge mediated supersymmetry breaking theories}},  {\em Phys. Lett. B} {\bf
  446} (1999) 28--36, [\href{http://arxiv.org/abs/hep-ph/9808401}{{\tt
  hep-ph/9808401}}].

\bibitem{Bolz2001}
M.~Bolz, A.~Brandenburg, and W.~Buchmuller, {\it {Thermal production of
  gravitinos}},  {\em Nucl. Phys. B} {\bf 606} (2001) 518--544,
  [\href{http://arxiv.org/abs/hep-ph/0012052}{{\tt hep-ph/0012052}}]. [Erratum:
  Nucl.Phys.B 790, 336--337 (2008)].

\end{thebibliography}\endgroup
\bibliographystyle{jhep}

\end{document}